\documentclass[a4paper, cleveref]{lipics-v2021}
\nolinenumbers
\pdfoutput=1
\usepackage[ruled,vlined]{algorithm2e}
\usepackage{bookmark}
\usepackage{booktabs}
\usepackage[subnum]{cases}
\usepackage{cite}
\usepackage{environ}
\usepackage[draft]{fixme}
\usepackage{paralist}
\usepackage{pgfplots}
\usepackage{pgfplotstable}
\usepackage{siunitx}
\usepackage{tabularray}
\usepackage{tikz}

\newcommand{\Omit}[1]{}

\fxsetup{theme=color}

\newcommand{\codefont}[0]{\sffamily}

\makeatletter
\lst@AddToHook{BoxUnsafe}{\everypar{}}%
\makeatother

\pgfplotsset{compat=1.17}

\usetikzlibrary{
  calc,
  chains,
  graphs,
  plotmarks,
  scopes,
  shapes,
}

\tikzset{
  artifact/.style=rounded corners,
  MU graph/.style={
    MU graph node/.style={ draw, execute at end node=\vphantom{Xy} },
    action node/.style={ MU graph node },
    data node/.style={ MU graph node, ellipse, },
    entry or exit node/.style 2 args={
      MU graph node,
      trapezium,
      trapezium angle=##1,
      node contents=##2,
    },
    entry node/.style={ entry or exit node={60}{entry} },
    exit node/.style={ entry or exit node={-60}{exit} },
    MU graph edge/.style={ ->, sloped },
    control edge/.style={ MU graph edge },
    data edge/.style={
      MU graph edge,
      dashed,
      edge node={node [auto, font=\small] {##1}}
    },
    definition edge/.style={ data edge=def },
    parameter edge/.style={ data edge=param ##1 },
  }
}

\pgfplotstableset{
  col sep=comma,
  columns/Description/.style=text column,
  every head row/.style={ before row=\toprule, after row=\midrule },
  every last row/.style={ after row=\bottomrule },
  multicolumn names,
  percentage column/.style={
    multiply by=100,
    numeric as string type,
    postproc cell content/.append style={
      /pgfplots/table/@cell content/.add={}{\,\%},
    },
  },
  text column/.style={ column type=l, multicolumn names=l, string type },
}

\UseTblrLibrary{booktabs}

\NewEnviron{GqlApi}{%
  \begin{booktabs}[expand=\BODY]{colspec=lX, rows={rowsep=.125\baselineskip}}
    \toprule
    Builder API & Description \\
    \midrule
    \BODY
    \bottomrule
  \end{booktabs}}

\Copyright{Rajdeep Mukherjee, Omer Tripp, Ben Liblit, and Michael Wilson}
\ccsdesc[100]{Theory of Computation, Program Analysis}

\title{Static Analysis for AWS Best Practices in Python Code}

\author{Rajdeep Mukherjee}{Amazon Web Services}{mukherr@amazon.com}{https://orcid.org/0000-0002-8179-4396}{}
\author{Omer Tripp}{Amazon Web Services}{omertrip@amazon.com}{https://orcid.org/0000-0002-2393-854X}{}
\author{Ben Liblit}{Amazon Web Services}{liblit@amazon.com}{https://orcid.org/0000-0002-2245-2839}{}
\author{Michael Wilson}{Amazon Web Services}{wilsonzy@amazon.com}{}{}
\authorrunning{R. Mukherjee, O. Tripp, B. Liblit, and M. Wilson}

\keywords{Python, Type Inference, AWS, Cloud, Best practices, Static Analysis}

\begin{document}

\maketitle

\begin{abstract}

Amazon Web Services (AWS) is a comprehensive and broadly adopted cloud provider,
offering over 200 fully featured services, including compute, database, storage, networking
and content delivery, machine learning, Internet of Things and many others.
AWS SDKs provide access to AWS services through API endpoints.
However, incorrect use of these APIs can lead
to code defects, crashes, performance issues, and other problems.
AWS best practices are a set of guidelines for correct and secure
use of these APIs to access cloud services, allowing conformant clients
to fully reap the benefits of cloud computing.

This paper presents automated static analysis rules, developed in the context of 
a commercial service for detection of code defects and security vulnerabilities,
 to identify deviations from AWS best practices
in Python applications that use the AWS SDK\@.
Such applications use the AWS SDK for Python, called ``Boto3'', to
access AWS cloud services. However, precise static analysis of Python applications 
that use cloud SDKs requires robust type inference for inferring the types of cloud
service clients.  The dynamic style of Boto3 APIs poses unique challenges for type resolution,
as does the interprocedural style in which service clients are used in practice.
In support of our best-practices goal, we present a layered strategy for type inference
that combines multiple type-resolution and tracking strategies in a staged manner:
(i) general-purpose type inference augmented by type annotations,
(ii) interprocedural dataflow analysis expressed in a domain-specific language, and
(iii) name-based resolution as a low-confidence fallback.
From our experiments across >3,000 popular Python GitHub repos that make use of the AWS SDK, 
our layered type inference system achieves 85\% precision and 100\% recall in inferring Boto3 clients in 
Python client code. 

Additionally,
we present a representative sample of eight AWS best-practice rules that detect a wide range of
issues including pagination, polling, and batch operations.  We have assessed the efficacy of these
rules based on real-world developer feedback. Developers have accepted more than 85\% of the recommendations
made by five out of eight Python rules, and almost 83\% of all recommendations.
\end{abstract}

\section{Introduction}
Python is a widely used language due to its simplified syntax and vast ecosystem
of libraries and frameworks, which save programmers time and effort.
Python is the language of choice for many developers in a wide range of application
domains, in particular in the cloud setting, including
Internet-of-Things (IoT), analytics, streaming, machine learning (ML), 
image processing, serverless applications, and many more.
Amazon Web Services (AWS) is a comprehensive and broadly adopted cloud provider.
Python is used extensively to build applications on top of the AWS cloud.  
This is done through the AWS SDK for Python, called Boto3, which mediates the interaction
between Python applications and AWS services.

Cloud SDKs enable access to rich functionality, spanning hundreds of services and thousands (if not more)
of API endpoints, but at the same time, the API contains subtleties and complexities that might lead 
to unexpected errors or performance issues.
AWS best practices are a set of guidelines that specify correct and secure usage of the AWS
cloud SDK\@.
It is imperative that applications follow AWS best practices to embrace the 
many benefits that the cloud offers. 

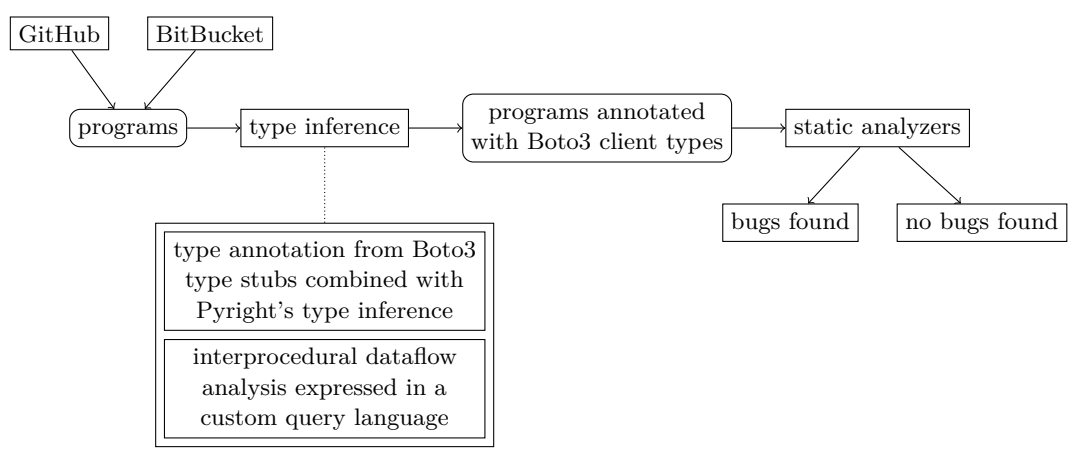
\begin{figure}
	\small
	\begin{tikzpicture}[every node/.style={draw, align=flush center}]

		{ [every on chain/.style={join=by ->}, node distance=.7cm, start chain=going right]
			\node (programs) [on chain, artifact] {\vphantom{X}programs} ;
			\node (inference) [on chain] {type inference\vphantom{y}} ;
			\node [on chain, artifact] {programs annotated \\ with Boto3 client types} ;
			\node (analyzers) [on chain] {static analyzers} ;
		}

		\node (github) [above=of programs, anchor=east, xshift=-.25cm] {GitHub} ;
		\node (bitbucket) [above=of programs, anchor=west, xshift=.25cm] {BitBucket} ;
		\draw [<-] (programs) edge (github) edge (bitbucket) ;

		\node (bugs found) [below=of analyzers, anchor=east, xshift=-.25cm] {bugs found} ;
		\node (no bugs found) [below=of analyzers, anchor=west, xshift=.25cm] {no bugs found} ;
		\draw [->] (analyzers) edge (bugs found) edge (no bugs found) ;

		\matrix (inference layers) [below=of inference, row sep=\pgfkeysvalueof{/pgf/inner ysep}, text width=4cm] {
			\node { type annotation from Boto3 type stubs combined with Pyright's type inference } ; \\
			\node { interprocedural dataflow analysis expressed in a custom query language } ; \\
		} ;
		\draw (inference layers) edge[densely dotted] (inference) ;
	\end{tikzpicture}
\caption{High-level overview of CodeGuru}
\label{codeguru}
\end{figure}

We report here on our experience in developing automated static analysis rules to enforce 
AWS best practices in Python applications. 
These rules are evaluated as part of a commercial cloud service, Amazon CodeGuru Reviewer (henceforth, CodeGuru),
\footnote{\url{https://docs.aws.amazon.com/codeguru/latest/reviewer-ug/welcome.html}} that runs 
static analysis on customer code to detect security vulnerabilities, optimization opportunities, and 
various other types of defects. The CodeGuru architecture is shown in \cref{codeguru}.

CodeGuru supports Java and Python, and integrates with different code hosting platforms including
GitHub and BitBucket. CodeGuru supports three code scanning modes:
\begin{itemize}
	\item Incremental: A code review is created automatically when a pull request is raised.
	\item Full: The entire codebase is analyzed. 
	\item CI/CD: The entire codebase is analyzed as part of CI/CD workflows.
\end{itemize}

\subsection{Importance of AWS Best Practices Rules}

In this section, we draw closely on real-world scenarios that were
experienced by CodeGuru customers. For confidentiality reasons, our description is of 
hypothetical scenarios, though these are inspired and motivated by real-world events.
We emphasize that all of these cases can be detected, and prevented, by applying static analysis 
to clients of the AWS SDK. Further, these scenarios are within the scope of the AWS best practices 
rules that we have developed, which alert the user to such risks already during code review.

We focus on large-scale operational failures. Previously, 
such failures would have to be discovered by manual inspection or testing. 
Such failures could result in critical issues such as race conditions leading to service outage 
or auto-ticketing errors; authorization and authentication errors; 
broken throttling mechanisms that impose unexpected loads on services, 
thereby leading to high latency or timeouts; missing or incorrect error handling, 
the impact being billing errors; and many other high-severity situations. 

Specifically:
\begin{itemize}
    \item A service that uses a paginated API that can retrieve a maximum of 1 MB of data 
or 1000 stream records, whichever comes first. The best practice for using paginated APIs is to 
check for additional results from the paginated API by iterating over the response object 
until no further pages can be retrieved. The service in question did not 
check for additional results from the paginated API, which eventually resulted in its 
availability rate dropping, and the service returning non-descriptive 5xx server errors to customers.
    \item A service used to ingest, buffer, and process streaming data in real-time 
utilizes the \lstinline{PutRecords} operation to send data into the stream for ingestion 
and processing. Each \lstinline{PutRecords} request can support up to 500 records. The
response includes an array of \lstinline{Record}s, including both successfully and 
unsuccessfully processed records. A single record failure does not block the processing 
of subsequent records. The service in question did not check for failed records
and erroneously treated all records as processed successfully, which 
resulted in data loss. The best practice in this case is to check the error code and message
on the returned \lstinline{Record}s, and re-attempt processing of failed records in a subsequent request.
    \item A fully managed messaging service for application-to-application communication 
provides topics for high-throughput, push-based, many-to-many messaging between 
distributed systems, microservices, and event-driven serverless applications. 
Consider an application utilizing this distributed messaging service 
that creates resources then immediately publishes to them. Due to the distributed
nature of the service, information might take some time to propagate, thus causing
some of the published messages to be dropped. The best practice is to sequence
topic creation/subscription and message publishing to mitigate such failures.
\end{itemize}
%
%
%

To give an idea of CodeGuru's throughput in a given week, we provide
lower-bound metrics. 
On average, CodeGuru analyzes $\gg10,000$
Pull Requests (PRs) containing $\gg1,000,000$ lines of code across $\gg100,000$ 
files, and provides $\gg1,000$ AWS best practices recommendations 
due to $\gg100$ different static analysis detectors.  


\subsection{Scope}

Though CodeGuru supports AWS best practices also for Java,
we focus on Python in particular given its dynamic nature and lack of static type annotations.
To enforce AWS best practices with high precision, it is essential to identify whether a given
function call is made against the AWS SDK, and if so, which service in particular is invoked. 
Unlike Java, where such information is provided through the type system, in Python 
this information is not available by default, which is the starting point for the research we report on 
in this paper.

We share details and results on several on-demand type resolution approaches,
as well as combinations thereof. We have experimented with the following core approaches:
\begin{itemize}
	\item[Type inference augmented with Boto3 type stubs:] Boto3 type stubs, 
    in combination with general purpose type inference, to resolve types when processing the Python AST\@.
	\item[Dataflow tracking]: On-demand interprocedural tracking, in both the forward and
	the backward directions, to check whether the receiver of a function call
	corresponds to a given AWS service.
	\item[API name based resolution:] An over-approximate yet lightweight strategy that simply
	checks whether the called function's name is compatible with a given
	AWS service's API\@.
\end{itemize}
We present not only the approaches themselves, and more advanced algorithms that combine 
these approaches, but also the underlying infrastructure that enabled us to implement these
approaches. Specifically, we provide technical details on the CodeGuru code representation and 
language for rule specification.

\subsection{Main Contributions}
 
To summarize, this paper makes the following principal contributions:
\begin{enumerate}
	\item Our main contribution is an on-demand type resolution strategy, which we demonstrate as effective in the case of Python clients of the AWS SDK\@.
	\item In support of the above-mentioned strategy, we present the Intermediate Representation (IR) and query language used by the CodeGuru service.
	\item We describe a representative sample of the AWS best practices rule suite running as part of the CodeGuru service.
	\item We share the details of the experiments we ran on 86,000 GitHub repositories, and real-world feedback we received from developers,
		to validate our type resolution strategy and the rules we built on top of it. 
\end{enumerate}

\subsection{Paper Structure} 

The rest of the paper is organized as follows. \Cref{Sec:related} sets the stage for our contributions in discussing related work. In \cref{Sec:background}, we establish the context for the specific problem that this paper addresses. We explain the design of the Boto3 library, and the information it provides on AWS services through type annotations and API models. Building on this background, we present in \cref{sec:motivating-examples} several code examples that motivate the need for advanced type inference, which is our main contribution in this paper. \Cref{Sec:representation,Se:GQL} lay down the technical infrastructure for our approach in describing, respectively, the code representation and query language that we use to express Python AWS best practices rules, and as part of these rules, also the data-flow tracking mechanisms for improved type inference. \Cref{inference} connects between earlier sections in describing the different capabilities that we have developed, based both on Boto3 type stubs and data-flow tracking, to resolve types in a wide range of scenarios. We then present a layered approach that integrates between the different type resolution strategies. With the supporting infrastructure and inference algorithm explained, we present in \cref{detectors} eight representative Python AWS best practices rules. \Cref{exp-result} is dedicated to our research hypotheses and experiments to validate the efficacy of our type inference algorithm. We conclude, and outline future research, in \cref{Sec:conclusion}.

\section{Related Work}\label{Sec:related}

Different approaches have been taken to infer Python type annotations, and formalize Python semantics more generally. We 
review approaches based on program analysis as well as machine learning, and compare these approaches with CodeGuru.

\subsection{Classical Program Analysis}

Widely used Python type checkers include mypy~\cite{MyPy},
Pyre~\cite{Pyre}, pytype~\cite{pytype}, and Pyright~\cite{Pyright}.
These tools rely on manual type annotations provided by developers,
augmented with varying forms of type inference.  However, retrofitting
type annotations onto large libraries or applications can be tedious
and error-prone.  Other prior work places more emphasis on static
analysis~\cite{DBLP:conf/cav/HassanUE018,DBLP:conf/dls/AnconaACM07,DBLP:conf/ecoop/MonatOM19,DBLP:conf/nfm/FromherzOM18,DBLP:conf/pepm/FritzH17,MITPythonThesis}
or dynamic analysis~\cite{DBLP:conf/dls/VitousekKSB14} to reduce
reliance on human-authored annotations.  Our initial search for
supporting infrastructure found that many published tools have failed
to keep up with recent Python releases, or omit support for key Python
features such as exceptions~\cite{DBLP:conf/pepm/FritzH17} or
recursion~\cite{DBLP:conf/ecoop/MonatOM19}.  We opted to use Pyright
as our baseline, as Pyright is both actively maintained and has a
rather advanced inference engine (see \cref{sec:pyright-features}).
In spite of these advantages, Pyright alone proved unsatisfactory for
our cloud application domain.  The details, as conveyed in
\cref{exp-result}, may serve to highlight challenges for other
developers of general-purpose type inference engines.

When writing type annotations, Python developers often focus on
function signatures: arguments and return values.  Some research tools
mirror this bias, such as
TypeWriter~\cite{DBLP:conf/sigsoft/PradelGL020}.  Xu et
al.~\cite{DBLP:conf/sigsoft/XuZCPX16} present a probabilistic type
inference system, but the accuracy of probabilistically inferred types
for Python variables is limited.  Our work requires accurate types for
variables, making these two approaches unsuitable.

Any attempt to statically analyze Python code must contend with the
intricacies of the Python language.  Notable efforts to formalize
Python semantics include those by
Smeding~\cite{Smeding:2009:PythonThesis}, Politz et
al.~\cite{DBLP:conf/oopsla/PolitzMMWPLCK13} and
Köhl~\cite{Koehl:2020:PythonThesis}.  Smeding's work predates Python
type annotations, while neither Politz et al. nor Köhl mention them
in any way.  These omissions are not surprising, as type annotations
have only limited effects on runtime behavior.  Thus, these
codifications of Python semantics offer little insight regarding the
type-inference challenges addressed here.  Our approach is neither
sound nor complete (see \cref{sec:mu-graph-construction}), so a
standard type-soundness theorem relating static types to runtime
semantics does not apply.

In the specialized domain of machine learning, where Python is perhaps
the most popular language, WALA Ariadne~\cite{DBLP:conf/pldi/DolbySAR18} 
analyzes Python specifically to infer the dimensions and types of tensors.  Like
Ariadne, our work is motivated by a specific application domain, and
even a specific framework: Ariadne focuses on machine learning using
TensorFlow~\cite{DBLP:conf/osdi/AbadiBCCDDDGIIK16}; CodeGuru
focuses on cloud computing using Boto3.  Ariadne's solution entails
both a custom type system and an analysis to infer it.  Our approach
builds upon standard Python types and type annotations.  While we
crafted our analysis strategy to match idiomatic Boto3 use, these
idioms are not exclusive to Boto3 client code.  Therefore our layered
approach may be more broadly applicable.

\subsection{Machine Learning}

PYInfer~\cite{DBLP:journals/corr/abs-2106-14316} uses deep learning to
generate type annotations for Python.  PYInfer fuses deep learning
with static analysis such as PySonar2 to infer types for variables as
well as function-level types in Python.  All of these techniques
either require labelled type annotations or employ a static analyzer
to generate the initial annotations from Python repositories in order
to train the deep neural network.  However, type resolution for Boto3
service clients is non-trivial due to the reasons mentioned above.

JSNice~\cite{DBLP:conf/popl/RaychevVK15},
DeepTyper~\cite{DBLP:conf/sigsoft/HellendoornBBA18}, and
LambdaNet~\cite{DBLP:conf/iclr/WeiGDD20} use deep learning to generate
type annotations for JavaScript and/or TypeScript.  LambdaNet's
authors note that TypeScript is an inviting target because ``plenty of
training data is available in terms of type-annotated programs.''  In
principle, similar strategies may be applicable to Python.  However,
it is unclear whether the available corpus of type-annotated Python
Boto3 client programs is large enough for effective training in
practice.

\section{Background on Boto3: the AWS SDK for Python} \label{Sec:background}

This section describes the AWS service clients in the AWS SDK for Python,
also called
``Boto3''.\footnote{\url{https://aws.amazon.com/sdk-for-python/}}

\subsection{Clients and Resources: Low- and High-Level APIs}
\label{sec:boto3-api-levels}

Boto3 has two distinct levels of APIs:

\begin{description}
\item[Client (or ``low-level'') APIs]
  provide one-to-one mappings to the underlying HTTP API operations.

\item[Resource APIs] hide explicit network calls but instead provide
  resource objects and collections to access attributes and perform
  actions. Resources represent an object-oriented interface to
  AWS. They provide a higher-level abstraction than the raw, low-level
  calls made by service clients.
\end{description}

A low-level service client can be created by passing the name of service as an argument to the
\lstinline{boto3.client} method.\footnote{\url{https://boto3.amazonaws.com/v1/documentation/api/latest/guide/clients.html}}
For example, the Python statement, \lstinline{s3_client = boto3.client('s3')},
creates a low-level client for the Amazon Simple Storage Service (S3).
Conversely, a service resource can be created by passing the name of service as an argument to the
SDK \lstinline{boto3.resource} method.\footnote{\url{https://boto3.amazonaws.com/v1/documentation/api/latest/guide/resources.html}}
For example, the Python statement, \lstinline{s3_client = boto3.resource('s3')},
creates an Amazon S3 service resource.
It is also possible to access the low-level client from an existing resource, as in:
\begin{quote}
\begin{lstlisting}
s3_resource = boto3.resource('s3')
s3_client = s3_resource.meta.client
\end{lstlisting}
\end{quote}
Alternatively, to use service resources, one can invoke the \lstinline{resource()} method of a \lstinline{Session} and pass in a service name.
For example, one can create an Amazon S3 service resource using:
\begin{quote}
\begin{lstlisting}
session = boto3.session.Session()
s3_resource = session.resource('s3')
\end{lstlisting}
\end{quote}
Service clients give access to service operations by calling methods
on a client. For example, suppose \lstinline{s3_client} is an S3
client.  Then one can create an S3 bucket, with the bucket name passed
via an argument, using:
\begin{quote}
\begin{lstlisting}
response = s3_client.create_bucket(Bucket=bucket_name)
\end{lstlisting}
\end{quote}

\subsection{Boto3 Type Stubs}
Boto3-stubs provides
full type annotations for Boto3.\footnote{\url{https://pypi.org/project/boto3-stubs/}} In particular, Boto3-stubs 
provides annotations for a \lstinline{Client} type, \lstinline{ServiceResource}, and 
\lstinline{Resource} type for each AWS service. It also provides annotations for a
\lstinline{Waiter} type, and a \lstinline{Paginator} type for each service. 
With help from Boto3-stubs, several Python type-checking tools can
discover types for multiple flavors of client 
construction calls such as \lstinline{boto3.client}, \lstinline{boto3.session}, 
\lstinline{session.client}, and \lstinline{session.session}. 

\subsection{API Specifications From Boto3} 
\label{sec:api-models}

Some of the AWS best practice rules that are presented in this paper use an external 
configuration that provides a specification of some service-specific fragment of the complete Boto3 API\@.
This specification includes an 
API name, type, the service name the API belongs to, and few other attributes that are 
relevant for the rule.  We refer to these external configurations as \emph{API specifications}.
One such example is presented in 
\cref{exp-result}.  API specifications are automatically extracted 
from Boto3 API models.\footnote{~\url{https://github.com/boto/boto3}}  
These API models have specific traits, such as, \emph{Pagination}, \emph{Batch}, 
\emph{Deprecated}, \emph{Waiters}, or \emph{mutual-exclusion}, which help 
determine the characteristics of the API.  We extract relevant API traits 
from API models across Boto3 services to construct the complete API specification
to enforce.  
These API specifications are then used by the best practice rules for analyzing 
client code.   

\section{Motivating Examples}
\label{sec:motivating-examples}

This section presents a few examples that motivate the need for
sophisticated type inference to recover the types of AWS service
clients in real-world Python applications.  The type annotations in
\cref{example1,example2} are obtained from Pyright with Boto3 type
stubs, which are on lines with the prefix
``\lstinline{#-->}''.

\begin{figure*}[tb]
  \begin{lstlisting}
import boto3

class S3(object):
  def __init__(self, **kwargs):
    self._client = None 
      
  @property
  def client(self)
    if self._client is None:
      self._client = boto3.client('s3')
   return self._client
 
  def M1(self):
    try:
      client = self._client
      #--> client: Optional[S3Client]
      # put lifecycle
      response = client.put_bucket_lifecycle(
          Bucket=test_bucket, LifecycleConfiguration=config)
      time.sleep(4)
      response = client.get_bucket_lifecycle(Bucket=test_bucket)
      assert response
      response = client.delete_bucket_lifecycle(Bucket=test_bucket)
    except CosServiceError as e:
      if e.get_status_code() < 500:
        raise e
  \end{lstlisting}
  \caption{Example of Python application code using Boto3}
  \label{example1}
\end{figure*}

\subparagraph{Example 1:}

Consider the Python code snippet in \cref{example1}.  The type
annotation for the variable, \lstinline{self._client}, in method
\lstinline{M1}, is shown in bold.  This type,
\lstinline{Optional[S3Client]}, suggests that \lstinline{self._client}
might be a Boto3 client for the Amazon S3 service.  The method
\lstinline{M1} performs a sequence of API calls on this S3 client
object:  \lstinline{put_bucket_lifecycle()}, then
\lstinline{get_bucket_lifecycle()}, then
\lstinline{delete_bucket_lifecycle()}.
If the type of \lstinline{self._client} could not be inferred in this
example, then it would have been difficult to guarantee the
correctness of the API calls invoked on that client. Hence, good type
inference for \lstinline{self._client} is crucial.

\begin{figure*}
\begin{lstlisting}
import boto3

class Skunky(object):
  def get_sns_client():
    return boto3.resource("sns")

  def M2():
    sns_arn = os.environ['PUBLISH']
    client = get_sns_client()
    #--> client: SNSServiceResource
    M3(client, topic, subscription)
    return client.Topic(sns_arn)  
    
  def M3(client, topic, subscription):
    topic = client.topic(topic)
    #--> (variable) client: Any
    subscription = topic.subscription(subscription)
    subscription.delete()
\end{lstlisting}
\caption{Example of a Python application code using Boto3}
\label{example2}
\end{figure*}

\subparagraph{Example 2:}

Consider the Python code snippet in \cref{example2}.  Here, the Boto3
client is returned by \lstinline{get_sns_client()}.  Its type is
\lstinline{SNSServiceResource}, marked in bold in method
\lstinline{M2}.  This type correctly identifies \lstinline{client} as
a client for the Amazon Simple Notification Service (SNS).  In
contrast to the previous example, \cref{example2} creates
\lstinline{client} using the \lstinline{boto3.resource()} API which gives an
object-oriented interface to
SNS.\footnote{\url{https://boto3.amazonaws.com/v1/documentation/api/latest/guide/resources.html}}.
The client flows into \lstinline{M3} via a function parameter.
\lstinline{M3} uses \lstinline{client} directly or indirectly to make
a sequence of API calls: \lstinline{topic()}, then
\lstinline{subscription()}, then \lstinline{delete()}.
Unfortunately, Pyright was unable to assign \lstinline{client} a
precise type, leaving it typed simply as the generic \lstinline{Any} inside
\lstinline{M3}.  Inference falls short here because Pyright cannot guarantee
that \lstinline{client} must \emph{always} be an \lstinline{SNSServiceResource}
in \emph{every} possible call to {M3}.  This is safe but, for our purposes, unfortunate:
an untyped \lstinline{client} cascades into
untyped \lstinline{topic} and \lstinline{subscription}, leaving us with
nothing useful to analyze for any of the API calls in \lstinline{M3}.

Type resolution of the variable \lstinline{client}
requires sophisticated type inference coupled with a domain-aware
preference for finding Boto3 clients wherever they \emph{might} arise
and be used for API interaction.  In this paper, we present a
technique that combines Pyright's type inference with a custom
interprocedural dataflow analysis to infer types in such cases.

Furthermore, these API names are exactly the same in Google's pubsub cloud
service,\footnote{\url{https://cloud.google.com/pubsub/docs}}
and AWS SNS service.  Our study shows that the names of some cloud
service APIs are exactly the same for cloud services from different
commercial cloud vendors (AWS, Google, Tencent, etc.).  Thus, precise
resolution of service clients' types is extremely important for
static analysis of Python applications that use these cloud SDKs.

CodeGuru is specifically developed to detect best-practice violations 
in Python application code that use the AWS Python SDK. Detection of 
APIs from other cloud vendors can be considered as a side-effect of 
the type inference strategy when it resorts to using the API name-based 
resolution as a fallback strategy. In such cases, our best-practice 
detectors can detect APIs that match names across different cloud 
vendors without any knowledge about the underlying type of cloud 
service clients. 

However, it would be possible to use the same program representation 
(see MU graph in \cref{Sec:representation}) and the Query
Language engine (see \cref{Se:GQL}) that are presented in this paper,
to codify best-practice recommendations for other cloud vendors as well. In terms of 
extensibility of our detectors to other cloud vendors, 
there are two aspects: (1) the type inference system, 
and (2) detector logic. The type inference system needs to be built for each 
specific cloud vendors and SDKs due to different styles and idiomatic ways 
of representing cloud service clients. The detector logic implementation 
can be reused for general purpose checkers such as missing pagination, 
deprecated APIs, batch APIs, mutually exclusive parameters, inefficient API chains, 
and few others. All these detectors depend on externally provided API 
specifications (see \cref{sec:api-models}) that are extracted from API models.
Reusability of the detector implementation depends on the availability of these 
API models for other cloud vendors.

\section{Program Representation}%
\label{Sec:representation}

Our analysis represents each program as a collection of per-function
graphs called \emph{MU graphs}.\footnote{``MU'' originally stood for
  ``misuse'', and is pronounced as the name of the Greek letter
  $\mu$.}  A MU graph roughly corresponds to a data-dependence graph
overlaid with a control-flow (not control-dependence) graph (CFG).  As
in prior work that used similar
representations~\cite{DBLP:conf/msr/AmannNNNM19,
  DBLP:journals/tse/AmannNNNM19}, we find this representation useful
for finding API misuse defects where both the data flowing into an
operation and the order of operations are important.

\subsection{MU-Graph Nodes}

MU graphs contain the following kinds of nodes:

\begin{description}
\item[Entry nodes] represent the start of a function's execution.
  Each MU graph has exactly one entry node.

\item[Exit nodes] represent the end of a function's execution.  Each
  MU graph has exactly one exit node.

\item[Control nodes] represent branched control flow among multiple
  possible successors, such as at a conditional statement or loop.

\item[Action nodes] represent individual execution steps, such as
  multiplying two values or calling a function.

\item[Data nodes] represent local variables, whether originally
  present in the source or added as temporaries during MU-graph
  construction.
\end{description}

Per-node metadata identifies specific uses of these general
categories.  For example, we distinguish a binary-multiplication
action from a function-call action, or an \lstinline{if}-statement
control node from a \lstinline{while}-statement control node.

Multiple assignments to the same local variable use multiple data
nodes, as in static single assignment (SSA) form.  $\phi$ action nodes
are added as needed to represent converging data flows, such as when
both branches of an \lstinline{if} statement modify the same variable.

\subsection{MU-Graph Edges}

MU-graph edges represent control and data flow:

\begin{description}
\item[Control edges] order execution among entry, exit, control, and
  action nodes.  No data node is ever the source or target of a
  control edge.  Thus, discarding all data nodes and non-control edges
  would reduce a MU graph to a traditional CFG\@.

\item[Data edges] represent movement of data among control and action
  nodes, and are further categorized as follows:

  \begin{description}
  \item[Condition edges] flow from data nodes into control nodes.  The
    source of a condition edge determines the control-flow path taken
    by the target of that edge.  For example, a condition edge flows
    from the value of an \lstinline{if} statement's predicate to the
    control node for the statement itself.

  \item[Definition edges] flow from an action to a data node defined
    by that action.  For example, a definition edge from a binary
    multiplication action to a data node $d$ indicates that $d$
    holds the result of that multiplication.

  \item[Parameter edges] flow from a data node into an argument
    position in an action node.  For example, a binary multiplication
    action is the target of two parameter edges, one for each operand.
    A function call action is the target of one parameter edge for
    each actual argument.

  \item[Receiver edges] flow from a data node into the receiver
    position in a call action node.  These are similar to parameter
    edges, but highlight the special role of the implicit
    \lstinline{self} or \lstinline[language=Java]{this} argument in
    method calls.

  \item[Callee edges] flow from a data node into the callee position
    in a call action node, showing what calculations identified the
    function to be called.  For example, in
    \lstinline{handlers[event]()}, 
    an indexing action to fetch \lstinline{handlers[event]} would
    define some temporary data node holding the function to call.  A
    callee edge would then flow from that data node to the call
    action.
  \end{description}
\end{description}

Edges carry additional metadata specific to their roles.  For example,
the two control edges that depart from an \lstinline{if} statement's
control node are marked to distinguish the true and false branches.
Multiple parameter edges leading to the same action node are ordered,
thereby distinguishing an action's first parameter from its second,
and so on.

\subsection{Overall Properties}

A key invariant in the MU representation is that data can only flow
from data nodes to control/action nodes, and vice versa.  Data edges
never connect pairs of data nodes directly, and every data edge has a
data node as exactly one of its endpoints.  Informally, each action
node receives zero or more data nodes as inputs (flowing across
parameter, receiver, or callee edges), and may provide an output that
flows across a definition edge into some other data node.  In a
compound expression like \lstinline{x + y * z}, the multiplication
action defines an anonymous data node, which in turn flows into the
addition action as its second argument.

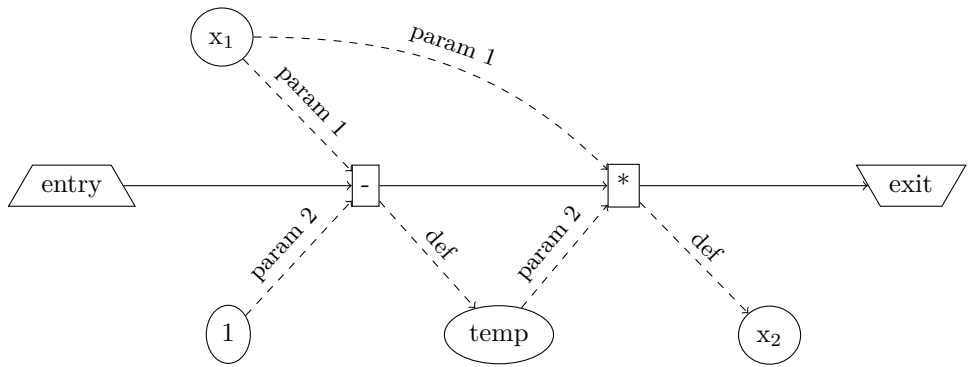
\begin{figure}
  \centering
  \begin{tikzpicture}[MU graph]
    { [
      start chain=going right,
      node distance=3cm,
      every on chain/.style={join=by control edge}
      ]
      \node (entry) [on chain,  entry node] {} ;
      \node (-) [on chain, action node] {-} ;
      \node (*) [on chain, action node] {*} ;
      \node (exit) [on chain, exit node] {} ;
    }

    { [every node/.style=data node, node distance=2cm]
      \node (x_1) [above left=of -] {$\text{x}_1$} ;
      \node (1) [below left=of -] {1} ;
      \node (x_2) [below right=of *] {$\text{x}_2$} ;
      \node (temp) at ($(1)!.5!(x_2)$) {temp} ;
    }

    \path
      (x_1) edge [parameter edge=1] (-)
      (1) edge [parameter edge=2] (-)
      (-) edge [definition edge] (temp)
      (x_1) edge [parameter edge=1, out=0] (*)
      (temp) edge [parameter edge=2] (*)
      (*) edge [definition edge] (x_2)
    ;
  \end{tikzpicture}
  \caption{The MU-graph representation of \lstinline{x *= x - 1}.
    Entry and exit nodes are trapezoidal; action nodes are
    rectangular; data nodes are elliptic.  Control edges are solid;
    data edges are dashed.\label{fig:mu-example}}
\end{figure}

\Cref{fig:mu-example} illustrates several MU-graph features in the
representation of \lstinline{x *= x - 1}, or equivalently %
\lstinline{x = x * (x - 1)}.  Solid control edges establish evaluation
order as in a CFG: subtraction before multiplication, each represented
as a rectangular action node.  Elliptic data nodes represent two
versions of \lstinline{x}: $\text{x}_1$ before the assignment and
$\text{x}_2$ after.  Additional data nodes represent the literal 1 and
a temporary value.  The initial value of \lstinline{x} ($\text{x}_1$)
is a parameter (operand) to both mathematical operations, and is
distinct from the final value of \lstinline{x} ($\text{x}_2$).  The
temporary data node is defined by the subtraction action and is also
a parameter to the multiplication action.  Notice that data and
non-data nodes strictly alternate along data paths:  data nodes
provide inputs to action or control nodes, and action nodes' outputs
define data nodes.

\subsection{Using Pyright for Best-Effort Graph Construction}
\label{sec:mu-graph-construction}

Pyright is ``a fast type checker meant for large Python source
bases.''~\cite{Pyright} Pyright is primarily used behind-the-scenes
to support Python IDEs, or as a command-line linter/checker.  However,
Pyright's sophisticated type inference and robust handling of
incomplete or incorrect programs make it ideally suited for our
purposes as well.

MU graph construction begins with a parsed abstract syntax tree (AST)
provided by Pyright.  We traverse the AST, synthesizing and combining
MU graph fragments in a roughly bottom-up manner.  For example, the MU
graph representation of an \lstinline{if} statement incorporates
smaller MU subgraphs representing the statement's conditional
expression, true branch, and optional false (\lstinline{else}) branch.

For data nodes, we also rely on Pyright to provide static type
information and name resolution.  Given Python's dynamic nature, these
are both best-effort.  Inferred static types can be imprecise, absent,
or wrong; names can be aliased or accessed covertly via reflection.
We attempt no alias analysis or points-to analysis beyond that
implicitly performed by Pyright's during type inference and name
resolution.  Types are available on data nodes that represent named
variables as well as those that represent intermediate subexpressions,
such as the ``temp'' node in \cref{fig:mu-example}, all subject to
practical limits on Pyright's ability to statically type Python code.

We flatten data node types to their string representations, such as
\lstinline{"int"} or \lstinline{"MyClass"} or %
\lstinline{"(int, str) -> tuple[int, str]"}.  Stringification discards
internal structure, but allows MU graphs to accommodate essentially
any type grammar, even from non-Python languages.  Types as strings
are also forgiving of incomplete programs:  we might know that a piece
of data is an instance of \lstinline{MyClass} even if we know nothing
about \lstinline{MyClass}'s internal structure or provenance.

The entire process of building MU graphs is best-effort, and proceeds
even when confronted with imports of missing modules, calls to unknown
functions, instantiations of unknown types, reads of unknown
variables, etc.  We represent each questionable operation as some
reasonable fallback (e.g., as an empty statement), and move on.
Python also contains syntactically ambiguous constructs, such as
overloaded operators or the myriad uses of ``\lstinline{.}''.  We
disambiguate these using types whenever possible, or heuristics when
necessary.  These approximations mean that we cannot, in general,
claim to be sound or complete.  However, these same approximations
allow us to provide a representation that is useful for many practical
applications where absolute guarantees are not required.

\subsection{From Functions to Programs}

The construction process described in \cref{sec:mu-graph-construction}
yields one MU graph for each function.  Beyond named functions
(\lstinline{def}), we also create a MU graph for each unnamed function
(\lstinline{lambda}).  For each top-level script, we create a
synthetic function that represents execution of that script's
top-level statements, and build a corresponding MU graph.

We organize these per-function MU graphs into larger assemblages that
reflect static program structure.  Each Python class contains a
dictionary of named methods; each script contains a dictionary of
named top-level classes and functions; and so on.  We do not build a
static call graph at this time, since not all downstream consumers of
MU graphs require one.  However, we organize and manage the MU graph
collection in such a way as to facilitate best-effort callee
resolution later, if needed.

\section{Query Language}\label{Se:GQL}

Working directly atop the MU representation in authoring analysis rules misses important reuse opportunities. We have therefore designed and implemented an API, dubbed the Golden Query Language (GQL), to enable encapsulation, optimization and reuse of a wide variety of analysis constructs. GQL is implemented as a Java library whose main interface with the analysis builder is the \lstinline{CustomRule} class. \lstinline{CustomRule} instances are created using the fluent builder pattern~\cite{DBLP:conf/ecoop/GammaHJV93}, where builder calls correspond to reasoning steps in the rule. A rule object can be evaluated at different scopes, from entire code bases to single functions. This is an important source of flexibility, which owes to the MU representation and its support for partial programs. (See \cref{sec:mu-graph-construction}.) Rule evaluation yields a \lstinline{RuleEvaluationResult} for every type and function that the rule visits, which includes rich information on whether, and if relevant where and how, rule evaluation has failed.

As an illustration of GQL syntax, we refer the reader to \cref{GQLRuleExample}, where a rule that identifies suboptimal use of the \lstinline{math.exp} function is shown. Here is a simple example of what the rule checks for:
\begin{quote}
\begin{lstlisting}
def foo():
  import math
  return math.exp(1e-10) - 1
\end{lstlisting}
\end{quote}

\begin{figure}
\begin{lstlisting}[language=Java]
CustomRule rule = new CustomRule.Builder()
  .withName("MathExp")
  .withComment("For small floats `x`, the subtraction in "
      + "`exp(x) - 1` can result in a loss of precision.")
  .withAllOf(
    b -> b
        .withMethodCallFilter(".*math\\.exp")
        .withDefinitionTransform()
        .as("MathExpResult"),
    b -> b
        .withConstantDataFilter("1")
        .as("ConstantOne")
  )
  .check()
  .withActionFilter("\\-")
  .withDirectDataFromIdFilter("MathExpResult")
  .withDirectDataFromIdFilter("ConstantOne")
  .build();
\end{lstlisting}
\caption{GQL rule for identifying suboptimal use of the \lstinline{math.exp} function.}
\label{GQLRuleExample}
\end{figure}

Rule definition begins by setting the rule's name and user-facing comment text. The following steps, up to the \lstinline{check} statement, are preconditions that the rule checks for. Specifically,
the \lstinline{withAllOf} statement ensures that all the subrules nested within it evaluate successfully, where these check for \lstinline{math.exp} calls as well as the presence of the constant value \lstinline{1}.
The matches are stored into variables (or IDs), to enable downstream reuse thereof, using the \lstinline{as} operation.
The actual check, or postcondition, is the rule section after the \lstinline{check} step.
It establishes whether there is a subtraction operation that the node defined by \lstinline{math.exp}, along with the constant \lstinline{1}, flow into directly (that is, without the mediation of any other action).

\subsection{Rule Evaluation}\label{Sec:RuleEvaluation}

In what follows, we use standard notation, $G=(V,E)$, when referring to MU graphs. Unless stated otherwise, the graphs we mention are specifically MU graphs.

As illustrated above, a GQL rule is an implication relation, $\mathit{pre} \implies \mathit{post}$. As such, rule evaluation is satisfied either when $\mathit{pre}$ is not satisfied or when both $\mathit{pre}$ and $\mathit{post}$ are satisfied. 
$\mathit{pre}$ and $\mathit{post}$ are both sequences $[\mathit{op}]$ of operations. 

An operation $\mathit{op} \colon \mathbb{P}(\mathcal{V})  \mapsto  \mathbb{P}(\mathcal{V})$ is a function whose domain and codomain are both node sets:
	$\mathcal{V} = \{n \colon \exists G=(V,E).\ n \in V\}$.
As an example, a filter operation that matches against calls to a function named ``\lstinline{foo}'' evaluates to \lstinline{foo} call nodes within the incoming node set, if any, or else $\emptyset$.

Given node $n$, let $G_n$ denote the graph containing $n$, and $G_n.V$ the complete set of nodes that $G_n$ contains. Operations $\mathit{op}$ satisfy the following two invariants:
\begin{compactenum}
	\item $\forall N \subseteq \mathcal{V}.\ op(N) \subseteq \bigcup_{n \in N} G_n.V$. That is, application of an operation to a node set $N$ cannot ``exceed'' the set of nodes due 
		to the graphs containing the nodes in $N$.
	\item $\mathit{op}(\emptyset) = \emptyset$. That is, application of an operation to the empty node set yields the empty node set.
\end{compactenum}

Given rule $r=[\mathit{op}_1,\ldots,\mathit{op}_k] \implies [\mathit{op}_{k+1},\ldots,\mathit{op}_n]$ and input graph $G=(V,E)$, we denote the node set flowing into $\mathit{op}_j$ as $\sigma_{j-1}$. The node sets 
are defined as follows:
\begin{quote} 
	$\sigma_i = 
		\begin{cases} 
			V   				& \mbox{if }i = 0 \\
			\emptyset 			& \mbox{if }i = k \wedge op_k(\sigma_{k-1}) = \emptyset \\
			V   				& \mbox{if }i = k \wedge op_k(\sigma_{k-1}) \neq \emptyset \\
			op_i(\sigma_{i-1}) 	& \mbox{otherwise}
		\end{cases}$
\end{quote}
Per the first case, precondition evaluation starts from the complete set of graph nodes ($V$). Per the second and third cases, the transition from precondition to postcondition is either trivial if the precondition has not been satisfied (second case), or --- analogously to precondition evaluation --- postcondition evaluation starts from $V$ (third case). Any other transition along the operation sequence is simply an application of the operation to its incoming node set. 

Rule evaluation is successful if and only if (i) a prefix of $\mathit{pre}$ evaluates to $\emptyset$ (in which case the precondition is not satisfied); or (ii) both $\mathit{pre}$ and $\mathit{post}$ evaluate to non-empty node sets (in which case the precondition and postcondition are both satisfied). 

To add color to the formal description so far, rule evaluation is essentially a process of matching against a pattern, or semantic property, where a non-empty node set is a \emph{match frontier} that feeds into the next reasoning step. Failure to maintain a non-empty match frontier means that the given (pre or post) condition is not satisfied by the input function.

\subsection{Rule Structure}

While our formal presentation above of GQL rules is as logical implication relationships, in practice a rule object has additional information and structure. A GQL rule consists of four sections, as follows:
(i) \emph{setup}: the rule's name, and the comment (or description) associated with the rule; (ii) \emph{function matcher}: a rule can optionally define criteria when to be evaluated, for example based on function name, attributes, annotations, containing type, parameter types, and so on; (iii) \emph{precondition}: the sequence of operations up to the \lstinline{check} builder step; and (iv) \emph{postcondition}: the sequence of operations following the \lstinline{check} builder step.

Since GQL rules follow the fluent builder pattern, there is risk that users would miss, misuse, or misorder rule constructs or sections. For example, the user might build a rule lacking a \lstinline{check} step; forget to set the rule's name; or try to apply incompatible filters in succession. To ensure rule integrity, we employ a hybrid solution that combines metadata contributed by operations with runtime checking. Operations expose a ``signature'', as explained in \cref{Sec:Operations}, such that improper compositions can be detected and localized ahead of rule evaluation.

\subsection{Language- and Domain-specific Rule Constructs}\label{GQLExtensibility}

Beyond the core GQL constructs, which are applicable across different programming languages and problem domains, there are reusable albeit language- or domain-specific constructs. As an example, constructs like \lstinline{withNamedArgumentsTransform} or \lstinline{withUnpackedArgumentsTransform} are useful for Python rules, but do not apply to Java. GQL enables such constructs to be organized into subclasses of \lstinline{CustomRule}, such as \lstinline{PythonCustomRule}, while containing \lstinline{CustomRule} to the core analysis constructs.

This approach has several important advantages. First, we avoid API bloat by distributing analysis constructs across more than just \lstinline{CustomRule}. Second, we avoid misuse errors due to a construct being used outside its intended context, for example a Python analysis construct used in a rule that targets Java programs. Finally, GQL extensions sometimes introduce dependencies. We have implemented, for example, a \lstinline{CustomRule} extension in the domain of data leaks, where some of the analysis constructs rely on an ML model to predict whether a given data access is retrieving sensitive information. These dependencies should not be forced on GQL users outside the given domain.

\subsection{GQL Operations}\label{Sec:Operations}

We now take a closer look at the different operations that comprise GQL rules. These divide into 4 categories, discussed below in turn. Beyond the description and examples in this section, we refer the reader to Appendix \ref{Sec:MoreGQLOps} for more examples of GQL operations.

For safety and fault localization, GQL requires that operations be annotated with their \emph{signature}, which states the types of nodes that they accept as input and yield as output. (See \cref{Sec:representation}.) The \lstinline{withReceiverTransform} operation, for example, accepts as input action (and more specifically, call) nodes, and outputs data nodes. If a user attempts to compose operations incorrectly, for example by routing the output of a \lstinline{withDataByNameFilter} operation to \lstinline{withReceiverTransform}, then GQL identifies the violation at runtime and generates a meaningful failure message that localizes it and explains why rule evaluation has been terminated. We are currently in the process of shifting the failure left to rule building time, and as a longer-term objective, compile time.

\subsubsection{Core Operations}

Core operations apply to all rules, regardless of their scope and logic. Some of the core operations, in particular \lstinline{check} and \lstinline{as}, have already been explained in the context of \cref{GQLRuleExample}. Additional core operations include the ability to reset the match frontier, interleave instrumentation (for example, for debugging or profiling purposes), read and write mutable auxiliary state, and so on. See \cref{Sec:rep-core-ops} for more examples.

\subsubsection{Filter Operations}

A filter operation $f$ satisfies the invariant:
	$\forall V \in \mathcal{V}.\ f(V) \subseteq V$.
That is, a filter operation selects a subset of the input node set. Its result cannot exceed the incoming set. 

GQL offers a wide selection of built-in filters. Beyond \lstinline{withActionFilter}, \lstinline{withMethodCallFilter}, \lstinline{withConstantDataFilter} and \lstinline{withDirectDataFromIdFilter} that are used in \cref{GQLRuleExample}, there are filters for matching against control structures, constants, actions with specific arguments (like constants or \lstinline{null}/\lstinline{None}), and so on. See \cref{Sec:rep-filter-ops} for several additional filter examples.
%


The GQL filter operations --- almost without exception --- are defined using a unary predicate ranging over nodes, and as such, filtering is done point-wise. As an example, \lstinline{withMethodCallFilter} is instantiated through a predicate that accepts action (and specifically, call) nodes where the callee matches the provided regex specification. One of the few exceptions to point-wise predicate application is the \lstinline{withMethodCallGroupFilter} operation, which evaluates sets of function calls and the relationships that hold between them (for example, if they all share the same receiver) against a user-provided specification.

A common practice with filter operations is to compose them, which enables refinement in pattern matching. The rule in \cref{GQLRuleExample} contains examples of that, like:
\begin{lstlisting}[language=Java]
.withActionFilter("\\-")
.withDirectDataFromIdFilter("MathExpResult")
.withDirectDataFromIdFilter("ConstantOne")
\end{lstlisting}
Here we first select subtraction actions then refine further by demanding unmediated incoming data flow from nodes mapped to variables \lstinline{"MathExpResult"} and \lstinline{"ConstantOne"}.

\subsubsection{Transform Operations}

Transform operations enable the transition from a given match frontier to another frontier that derives from it. For example, a frontier that consists of function calls can be transformed to the respective arguments or receivers, or the values defined by the calls, as illustrated with \lstinline{withDefinitionTransform} in \cref{GQLRuleExample}.

There are many built-in GQL transform operations. Examples include \lstinline{withArgumentsTransform}, which transforms an action node to its respective arguments; \lstinline{withControlDependenciesTransform}, which transforms a node to its set of control dependencies; \lstinline{withDataDependenciesTransform} (\textit{resp.} \lstinline{withDataDependentsTransform}), which transforms a node to its set of (transitive) data dependencies (\textit{resp.} dependents); and \lstinline{withReceiverTransform},  which transforms a call node to the receiver (if available). See \cref{Sec:rep-transform-ops}.

\subsubsection{Second-order Operations}

Logical structures and operators are necessary to express certain rule logic in a precise and concise manner. As a simple example, the user may wish to check if a given function call \lstinline{"zoo"} has a receiver of type either \lstinline{Foo} or \lstinline{Bar}. Another use case, illustrated in \cref{GQLRuleExample}, is the need to check that several conditions are all met through \lstinline{withAllOf}.

To enable such control and logical structures, GQL exposes second-order operations. These are operations that are themselves parameterized by one or more rules, which we refer to as \emph{subrules}. 

As an illustration, here is the GQL syntax for the above example:
\begin{lstlisting}[language=Java]
.withMethodCallFilter("zoo")
.withOneOf(
  b -> b.withReceiverByTypeFilter("Foo"),
  b -> b.withReceiverByTypeFilter("Bar"))
\end{lstlisting}
The \lstinline{withOneOf} construct evaluates to the first subrule that yields a non-empty result, or else it evaluates to $\emptyset$.

Most of the computational complexity and analysis power of GQL is attributable to this category of operations, which includes support for interprocedural analysis through \lstinline{withInterproceduralMatch}, described in \cref{Se:Interproc}, and variants thereof like \lstinline{withInterproceduralDataDependenciesTransform} or \lstinline{withInterproceduralDataDependentsTransform}; different forms of aggregation of subrule evaluation results, like \lstinline{withOneOf}, \lstinline{withAllOf} or \lstinline{withAnyOf} (collecting all non-empty subrule evaluation results); \lstinline{withLanguageSpecific} (which enable language-specific subrule logic, for example to resolve types differently in Java versus Python); and so on. For more examples of second-order operations, see \cref{Sec:rep-2ndorder-ops}.

\subsection{Interprocedural Analysis}\label{Se:Interproc}

As noted above, GQL provides the ability to perform interprocedural analysis through the \lstinline{withInterproceduralMatch} construct and several specializations thereof. The underlying call-graph representation resolves call sites on demand, per the CHA call-graph construction algorithm~\cite{DBLP:journals/toplas/GroveC01}, based on the (i) name of the callee, (ii) number of arguments, and (iii) argument types. Though the CHA algorithm is known to be
imprecise~\cite{DBLP:conf/oopsla/BaconS96}, we have rarely seen cases where that was the cause of imprecision in GQL rule evaluation. We hypothesize that this is because (i) interprocedural analysis is run at file or package scope, but not beyond, so there is less room for error, plus (ii) imprecision in interprocedural analysis is potentially mitigated by other rule steps.

A pseudocode description of the GQL support for interprocedural tracking is shown in \cref{Alg:interproc}. Interprocedural tracking takes a subrule $r$, along with a pair $\langle G_0, mr_0 \rangle$ of seeding function graph and matched nodes within it, and a specification whether to track forward or backward. The algorithm computes a fixpoint solution mapping visited functions to matched nodes therein.

At each iteration of the fixpoint loop, the rule is applied to the current worklist item $\langle G, mr \rangle$, computing additional matches $mr^\prime$ beyond the seeding matches $mr$. In addition, $mr$ and $mr^\prime$ are used to derive ``summaries'' at call sites $c$ dispatching to $G$, which is shown in \cref{Alg:interproc} as the $propagateAtCallerSites$ step. In the forward direction, we check whether there are matched $c$ arguments that are compatible with $mr$, and if so, we track additional arguments and/or the node defined by $c$, if exists, if these are supported by $mr^\prime$ (that is, if $mr^\prime$ contains formal parameters of $G$ and/or nodes returned by $G$). In the backward direction, we check whether $mr$ contains nodes returned by $G$ and $mr^\prime$ contains formal parameters, and if so, then we start tracking corresponding $c$ arguments.

We then proceed to additional propagation steps. In the forward direction, we extend the worklist by (i) propagating from call arguments to respective callee parameters, and (ii) propagating from return statements to nodes defined by compatible call sites. In the backward direction, we extend the worklist by (i) propagating from formal parameters to compatible call-site arguments, and (ii) propagating from nodes defined by call sites to returned nodes of compatible callee.

\begin{algorithm}
 \KwData{$\langle G_{0}, mr_{0} \rangle$: function graph and matched nodes therein}
  \KwData{$rule$: intraprocedural matching rule}
 \KwData{$isFwdElseBack$: analysis direction}
 \KwResult{$\{ \ldots, G \mapsto mr, \ldots \}$: fixpoint matching solution}
 $worklist = \{ \langle G_{0}, mr_{0} \rangle \}$\;
 $result = \emptyset$\;
 \While{$\neg worklist.isEmpty()$}{
  $\langle G, mr \rangle$ = $worklist.selectAny()$\;
  \If{already processed $\langle G, mr \rangle$} {
    $continue$\;
  }
  $mr^\prime = rule.evaluate(G, mr)$\;
  $result = result[G \mapsto mr \cup mr^\prime]$\;
  $\{ \ldots, G_{clr_i}, \ldots \} = resolveCallers(G)$\;
  \ForEach{caller $G_{clr_i}$}{
    $propagateAtCallerSites(G_{clr_i}, G, mr, mr^\prime, isFwdElseBack)$
  }
  \eIf{$isFwdElseBack$}{
    \ForEach{call $c$ with matched args in $mr^\prime$}{ 
        $\{ \ldots, G_{tgt_i}, \ldots \} = resolveCallees(c)$\; 
        \ForEach{callee $G_{tgt_i}$}{
          $mr_{tgt_i} = argsToFormals(c, mr^\prime, G_{tgt_i})$\;
          $worklist.add(\langle G_{tgt_i}, mr_{tgt_i} \rangle)$\;
        }
     }
     \ForEach{return statement $ret$ with matched arg in $mr^\prime$}{  
        \ForEach{caller $G_{clr_i}$}{
          $mr_{defs} = toCallerSiteDefs(ret, G, G_{clr_i})$\;
          $worklist.add(\langle G_{clr_i}, mr_{defs} \rangle)$\;
        }
     }
   }{
   \ForEach{parameter $p$ in $mr^\prime$}{  
        \ForEach{caller $G_{clr_i}$}{
          $mr_{clr_i} = formalToCallerSiteArgs(p, G, G_{clr_i})$\;
          $worklist.add(\langle G_{clr_i}, mr_{clr_i} \rangle)$\;
        }
     }
     \ForEach{call $c$ whose definition is in $mr^\prime$}{  
        $\{ \ldots, G_{tgt_i}, \ldots \} = resolveCallees(c)$\;
        \ForEach{callee $G_{tgt_i}$}{
         $mr_{tgt_i} = returned(G_{tgt_i})$\;
         $worklist.add(\langle G_{tgt_i}, mr_{tgt_i} \rangle)$\;
        }
     }
  }
  \Return{$result$}\;
 }
 \caption{Interprocedural matching algorithm}\label{Alg:interproc}
 
\end{algorithm}

\subsection{Dataflow Analysis}

Beyond its interprocedural capabilities, GQL also has built-in support for several flavors of dataflow  analysis, including slicing and taint tracking.\footnote{
	GQL additionally features finite state machine (FSM) and typestate analysis, though these involve not just dataflow  but also control-flow reasoning. These capabilities 
	are not consumed by the rules that we discuss later in the paper, so we suffice by noting them here.
} These build directly on top of the data edges exposed by the MU representation, in conjunction with the interprocedural matching algorithm described above. 

The main feature that the GQL dataflow  analysis provides beyond a standard fixpoint algorithm over the dataflow  relation is the ability to specify matchers on graph edges to tag them with unique roles: \emph{passthrough} (data flows across the call site), \emph{blocking} (an edge being either a sanitizer or a validator), \emph{side effecting} (data flows into the receiver of a call), or \emph{reading} (data flows from the receiver to the definition). The user-provided specification is then enforced as part of the fixpoint algorithm.

\section{Type Inference for Boto3 Clients}~\label{inference}

As explained in \cref{sec:boto3-api-levels},
a Python AWS application creates an AWS service client by passing the
name of the service as an argument to one of two distinct levels of
APIs.  The use of these multiple API flavors, the
interactions between them, and the use of strings as service
selectors, all pose challenges for type inference.

Regardless of which API is used, AWS service clients are ultimately
just data values.  Like any other data, service clients can be stored
in class variables, assigned into global variables, returned from
functions, and so on.  Code might use a service client locally within
a single function or globally within or even across the files that
comprise the complete application.  The complexity of these
\emph{definition--use chains} (DU chains) further complicates type
inference.

In this section, we present different type inference strategies that
can be used in this challenging application domain.

\subsection{Pyright's Type Inference With Boto3-Stubs}
\label{sec:pyright-features}

Pyright supports type inference for function return values, instance
variables, class variables, local variables, and global variables.
Pyright's inference engine uses several advanced type inference
techniques, such as:

\begin{itemize}
\item a flexible model of ``type assignability'';

\item inferred types for \lstinline{self} and \lstinline{cls};

\item parameterized generic types, including both polymorphic
  container types as well as optional types;

\item union types representing arbitrary sets of possible types;

\item overloaded function types as a special case of union types for
  \emph{ad hoc} polymorphic functions;

\item literal types, such \lstinline{Literal["str"]} as a subtype of
  \lstinline{str} that represents only the string literal
  \lstinline{"s3"}.

\item type narrowing based on code flow, which naturally makes
  Pyright's types flow-sensitive;

\item type guards that recover implicit type constraints from a wide
  variety of common Python idioms;

\item constrained type variables and conditional types; and

\item bidirectional type inference that infers the ``expected type''
  of the right-hand side of an assignment if the left-hand side
  already has a known type.
\end{itemize}

A full discussion of these capabilities is outside of the scope of
this paper, and in any case Pyright is not our contribution.  We treat
Pyright's type inference engine as a powerful, featureful, but opaque
black box.

If Pyright cannot infer the type of some symbol, then that symbol's
type is set to \lstinline{Any}.  This fallback type is a useful
warning marker that lets inference consumers (such as CodeGuru)
recognize cases where Pyright type inference fell short.

Type inference can incur significant computation overhead for large
code bases.  Also, Pyright cannot always infer correct types without
some outside help.  Hence, type annotations are a practical
requirement for building a robust type inference system.  We use
third-party type stubs, called \emph{Boto3-stubs}~\cite{Boto3-stubs},
that provide full type annotations for Boto3.  Pyright
ingests type annotations provided by Boto3-stubs to further enhance
and constrain its type inference.

\subparagraph{Examples of Pyright's Type Inference:}

\Cref{example1,example2} give the variable- or function-level type
annotations from Pyright with Boto3-stubs (denoted by the prefix
``\lstinline{#-->}''), where
Boto3 clients are stored in instance fields, or returned from methods,
respectively.  However, in \cref{input-parameter-not-annotated},
Pyright fails to infer a precise type for \lstinline{s3_client} in the
method \lstinline{load_df_from_s3}, instead giving it the fallback
\lstinline{Any} type.

\subsection{Type Inference Using Custom Dataflow Rules}

As an alternative to Pyright, we have used GQL to implement custom
inference rules based on dataflow analysis.  These rules do not
provide universal, generic type inference.  Instead, they focus on
idiomatic, interprocedural Boto3 usage patterns that Pyright's
general-purpose engine fails to address.  There are a total of 
ten GQL-based custom dataflow rules, among which only one is 
intraprocedural rule and rest nine are interprocedural rules.
For illustration purpose, we select few representative interprocedural
GQL rules that have low to medium complexity (in terms of number of 
operations in the rules) and that performs dataflow analysis at file-scope 
or package-scope. 

\subsubsection{Representative Examples of Interprocedural Rules}
\label{sec:rule-examples}

Each GQL rule in
\crefrange{fig:forward-interprocedural}{fig:backward-interprocedural-field}
implements some form of interprocedural dataflow analysis.  Each
operates on a function graph and matching API nodes along with the
receiver nodes of calls to the corresponding APIs.  For example, in
\cref{input-parameter-not-annotated}, one relevant API node is
\lstinline{get_object}, for which the corresponding receiver node is
\lstinline{s3_client}.  Our strategy for resolving call actions to
callees is name-based:  we match the name of the API entry point
(callee) in the code against API specifications that are extracted 
from Boto3.  

\newcommand{\rulefigure}[2]{%
  \begin{figure*}
    \lstinputlisting[language=Java]{#1.java}
    \caption{Rule example using #2\label{fig:#1}}
  \end{figure*}}

\rulefigure{forward-interprocedural}{forward, interprocedural
  dataflow}

\Cref{fig:forward-interprocedural} shows one such rule.  The scope of
this rule's interprocedural match operation is
\lstinline[language=Java]{FILE_FORWARD_REACHABLE}, %
which directs GQL to track dataflow forward using a ``data
dependents'' transform operation that transforms from incoming nodes
to nodes that are data dependent on them, including in other
functions.  The result of this interprocedural tracking is then
checked to determine if it matches one of the known flavors of Boto3
clients (low-level or object-oriented), by calling the utility methods
inside the \lstinline[language=Java]{withOneOf} operation.

\rulefigure{backward-interprocedural}{backward, interprocedural
  dataflow}

The rule in \cref{fig:backward-interprocedural} implements
interprocedural backward dataflow analysis, complementary to the
forward analysis of \cref{fig:forward-interprocedural}.  For the
backward version, tracking is specified as
\lstinline[language=Java]{FILE_BACKWARD_REACHABLE}. %
This scope directs the interprocedural analysis to perform backward
dataflow tracking using a ``data dependencies'' transformer that
transforms from incoming nodes to nodes that are data dependent on
them, including in other functions.  Similar to the previous rule,
this rule's \lstinline[language=Java]{withOneOf} clause then checks
whether the result of backward interprocedural tracking matches one of
the known flavors of Boto3 clients.

\rulefigure{backward-interprocedural-field}{backward, interprocedural
  dataflow for receivers in instance fields}

\Cref{fig:backward-interprocedural-field} gives an interprocedural
rule for tracking Boto3 clients when the client is initially stored in
an instance field by a constructor (``\lstinline{__init__}''), then
later retrieved from the same field and used as the receiver node of a
matching API call.  This rule uses a
\lstinline[language=Java]{withNodeTransform} %
operator that transforms all field accesses within the package scope
to only those nodes that match with the receiver node.  An
interprocedural query on this matching set retrieves all data nodes
that are backward-reachable within the package scope. The resultant
set of nodes is passed to the
\lstinline[language=Java]{withDefinedTransform} that transforms from
incoming data nodes to their respective defining actions.  Lastly,
this rule checks that the resultant set of action nodes match one of
the known flavors of Boto3 clients.

\subsubsection{Example of Type Inference Using Custom Dataflow Rules}

\begin{figure*}
  \begin{lstlisting}
def write_df_to_s3_location(file_path, bucket_name, \
    metadata_material_set, sep=None):
  s3_client = create_s3_client()
  #--> s3_client: S3Client
  df = load_df_from_s3(s3_client, bucket=bucket_name, path="")
  #--> df: DataFrame
  s3_client.put_object(Body=file_path, Bucket=bucket_name)

def create_s3_client():
  return Boto3.client("s3")
  #--> create_s3_client: () -> S3Client

def load_df_from_s3(s3_client, bucket, path):
  #--> load_df_from_s3: (s3_client, bucket, path) -> DataFrame
  raw_date = None
  try:
    raw_data = s3_client.get_object(
      Bucket=bucket_name, Key=object_path)
    #--> s3_client: Any
    #--> s3_client: Any
  except ClientError as e:
    logging.info("Bucket: {}, error {}", bucket_name, str(e))
  io_data = StringIO(raw_data)
  df = pd.read_csv(io_data)
  return df
  \end{lstlisting}
  \caption{Type annotation for AWS client passed by input parameter}
  \label{input-parameter-not-annotated}
\end{figure*}

\Cref{input-parameter-not-annotated} shows a Python code snippet with
variable- and function-level type annotations from Pyright.  The type
of \lstinline{s3_client} in the method
\lstinline{write_df_to_s3_location} is correctly inferred as
\lstinline{S3Client}: an Amazon S3 service client. This client is
passed via input parameter to the method
\lstinline{load_df_from_s3}. In absence of the type annotation for the
input parameter, Pyright could not infer the type of
\lstinline{s3_client} (denoted by \lstinline{Any}), inside the method
\lstinline{load_df_from_s3}.

However, another of our custom dataflow rules can resolve the type of
\lstinline{s3_client} in method \lstinline{load_df_from_s3}.  The
applicable rule starts from a matching API node,
\lstinline{s3_client.get_object}, where the type of the receiver node
\lstinline{s3_client} needs to be determined.  Recall that the
matching API node is obtained by matching the name of the API against 
the API specification extracted from Boto3.  Starting from a
matching API node, the rule uses a ``parameter transform'' operation
that transforms incoming nodes to the parameters of the respective
functions.  This rule then uses a ``backward data dependencies''
transform that transforms from incoming nodes to their data
dependencies, including in other functions.  The rule's result
includes the node \lstinline{s3_client} in the method
\lstinline{write_df_to_s3_location}, whose type is already known to be
\lstinline{S3Client}.  It is worth noting that the type of
\lstinline{s3_client} could also be inferred by a stand-alone custom
dataflow rule (in absence of type annotations from Pyright).  However,
the rule specification would be more complex.  We prefer to augment
Pyright's capabilities rather than replace them.

\subsection{Layered Type Inference}

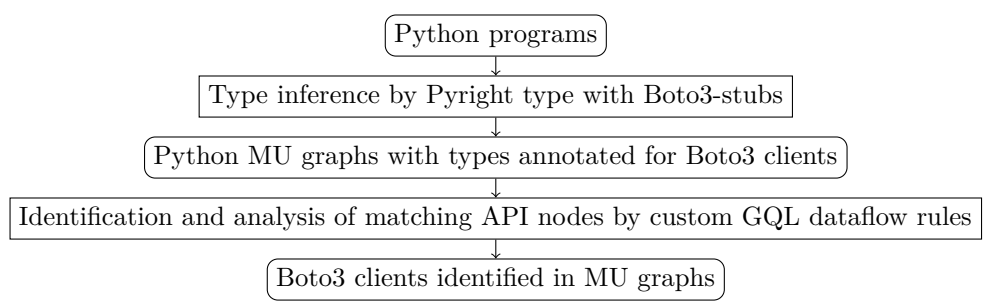
\begin{figure*}
  \centering
  \begin{tikzpicture}
    [
      every node/.style={ align=center, draw, on chain },
      every on chain/.style={ join=by -> },
      node distance=.25cm,
      start chain=going below,
    ]
    \node [artifact] {Python programs} ;
    \node {Type inference by Pyright type with Boto3-stubs} ;
    \node [artifact] {Python MU graphs with types annotated for Boto3 clients} ;
    \node {Identification and analysis of matching API nodes by custom GQL dataflow rules} ;
    \node [artifact] {Boto3 clients identified in MU graphs} ;
  \end{tikzpicture}
  \caption{Layered type inference\label{type-inference}}
\end{figure*}

The example in \cref{input-parameter-not-annotated} shows that a
hybrid approach for type inference can combine custom dataflow rules
with Pyright's type inference to resolve types that Pyright cannot
resolve by itself.  Each of these type inference approaches have
complementary strengths.  This quality suggests a layered approach for
type inference that combines these strategies in a staged manner, as
shown in \cref{type-inference}.

Our layered approach first uses Pyright's type inference with Boto3
stubs to infer type annotations for at least some Boto3 clients.  Per
\cref{sec:mu-graph-construction}, data nodes in MU graphs carry type
metadata reflecting Pyright's inference results.  If the type of an
API call of interest is already known, then that may be sufficient to
recognize that the API belongs to Boto3.  If the type of the API call
of interest is unknown, then our layered approach deploys custom
dataflow rules to infer client types.  \Cref{exp-result} presents our
empirical evaluation of the strengths and limitations of this layered
approach.

\section{AWS Best Practices Rules}~\label{detectors}
In this section, we describe a representative sample of eight rules that 
detect different types of defects related to usage of the Boto3 API\@.
These rules cover approximately 200 public-facing AWS services. 
All Python AWS best practices rules (as well as most other CodeGuru rules) 
are implemented atop GQL (see \cref{Se:GQL}), and follow the
same rule evaluation mechanism that is discussed in \cref{GQLRuleExample}.
Of the eight rules discussed in this section, we focus in particular on two rules ---
concerning pagination and batchable APIs --- to enable thorough discussion
of rule syntax and sample detections.
 
\subsection{Detecting Misuse of Paginated APIs}

The pagination trait is implemented by over 1,000 APIs belonging to >150
AWS services. This trait is commonly used when the result set due to a 
query is too large to fit within a single response. For the complete set 
of results, a pagination token is used to perform iterative requests and 
receive the response in parts.  Developers who are not aware of this trait 
might mistakenly suffice with a single request/response result, as 
illustrated in \cref{missing_pagination}.

\begin{figure*}
\begin{lstlisting}
def sync_ddb_table(source_ddb, destination_ddb):
    response = source_ddb.scan(TableName="table1")
    for item in response['Items']:
        destination_ddb.put_item(TableName="table2", Item=item)
\end{lstlisting}
\caption{Non-compliant Pagination Example}
\label{missing_pagination}
\end{figure*}

Here the \lstinline{scan} call is used to read items from an Amazon DynamoDB table, where
\lstinline{put_item} saves those items to another DynamoDB table.
The \lstinline{scan} API implements
the pagination trait. However, the code neglects to check for additional results
beyond the initial batch, which is clearly wrong.
Our pagination rule detects the missing pagination in this example, and generates a recommendation to
iterate on the complete result set through the \lstinline{LastEvaluatedKey} token
available through \lstinline{response}. A compliant version of the code, consistent
with this recommendation is shown in \cref{correct_pagination}.

\begin{figure*}
\begin{lstlisting}
def sync_ddb_table(source_ddb, destination_ddb):
    response = source_ddb.scan(TableName=="table1")
    for item in response['Items']:
        destination_ddb.put_item(TableName="table2", Item=item)
    # Keeps scanning util LastEvaluatedKey is null
    while "LastEvaluatedKey" in response:
        response = source_ddb.scan(
            TableName="table1",
            ExclusiveStartKey=response["LastEvaluatedKey"]
        )
        for item in response['Items']:
            destination_ddb.put_item(TableName="table2", Item=item)
\end{lstlisting}
\caption{Correct Pagination Example}
\label{correct_pagination}
\end{figure*}

\subsection{Error Handling for Batch Operations}

More than 20 AWS services expose batch APIs, which enable bulk request processing. 
Batch operations can succeed without throwing an exception even if
processing fails for some items. Therefore, a recommended best practice is to
explicitly check for failures in the response due to the batch API call.

\begin{figure}
\includegraphics[width=1.10\textwidth]{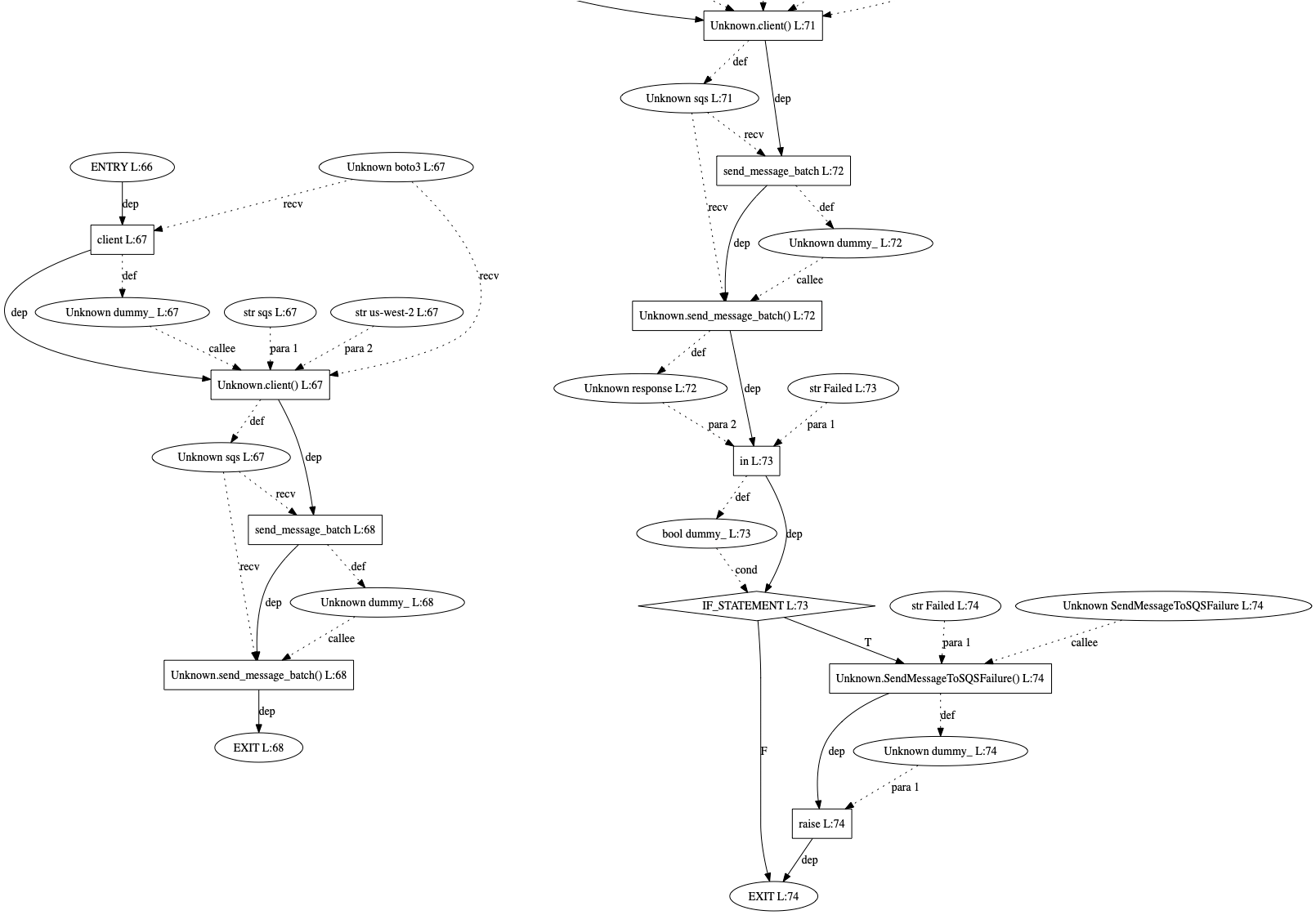}
\vspace*{-3mm}
\caption{Complete MU Graph (left) for the non-compliant code and 
partial MU Graph (right) for the compliant code of
\cref{incorrect_batch,correct_batch}, respectively}
\label{mu-graphs}
\end{figure}

We illustrate incorrect and correct usages of batch APIs in 
\Cref{incorrect_batch,correct_batch}, respectively. The MU representation for these programs is provided in \cref{mu-graphs}.

\begin{figure*}
\begin{lstlisting}
def noncompliant():
  sqs = boto3.client('sqs', 'us-west-2')
  sqs.send_message_batch()
\end{lstlisting}
\caption{Incorrect Error handling for Batch Operation example}
\label{incorrect_batch}
\end{figure*}
\begin{figure*}
\begin{lstlisting}
def compliant():
  sqs = boto3.client('sqs', 'us-west-2')
  response = sqs.send_message_batch()
  if "Failed" in response:
    raise SendMessageToSQSFailure("Failed")
\end{lstlisting}
\caption{Correct Error handling for Batch Operation example}
\label{correct_batch}
\end{figure*}

The rule for detection of batch operations where failures are not checked is 
shown in \cref{Fig:batchOpsRule}. Like many other CodeGuru rules,
in particular in the AWS best practices category, this rule is parameterized by
 a configuration. (See \cref{exp-result} for an example.)
 
The rule's precondition searches for batch API calls per the configuration, then
transforms from the calls to their respective receivers, which are stored into
variable \lstinline{PYTHON_AWS_CLIENT_RECEIVER}. Backward propagation,
in an attempt to relate these receiver nodes to applicable Boto3 services,  
then takes place through the \lstinline{getBoto3} call.

The postcondition loads the batch API call, stored as variable \lstinline{BATCH_API_CALL},
then checks whether the result of the call is ignored through \lstinline{withOutputIgnoredFilter}. This
filter checks whether the call node(s) flowing into it define(s) a node that has no outgoing edges.

\begin{figure}
\begin{lstlisting}[language=Java]
PythonCustomRule.Builder()
    .withName(ERROR_HANDLING_BATCH_OPERATION_RULE)
    .withComment(ERROR_HANDLING_BATCH_OPERATION_RULE_COMMENT)
    .withMethodCallFilter(config.api)
    .as(BATCH_API_CALL)
    .withReceiverTransform()
    .as(PYTHON_AWS_CLIENT_RECEIVER)
    .reset()
    .withClosure(
        /* Pre-condition: Match that the type of API is a Boto3 client */
        b ->
            getBoto3Client(
                (PythonCustomRule) b,
                BotoServiceName.fromString(config.serviceId),
                PYTHON_AWS_CLIENT_RECEIVER))
    /* CHECK */
    .check()
    /* Post-condition: Check that the output of Boto3 API is ignored */
    .withId(BATCH_API_CALL)
    .withOutputIgnoredFilter()
    .build();
\end{lstlisting}
\caption{Rule to check for batch API calls sans failure checking.}
\label{Fig:batchOpsRule}
\end{figure}

\subsection{Other Representative Rules}

We now switch to additional rules in the AWS best practices category, and provide
an explanation of what they each check for.

\begin{itemize}
\item[Use waiters in place of polling API:] 
Waiters are utility methods that make it easy to wait for a resource 
to transition into a desired state by abstracting out the polling 
logic into a simple API call. The waiters interface provides 
a custom delay strategy to control the sleep time between retries, as well as a 
custom condition on whether polling of a resource should be retried. 
Our rules detect code that appears to be waiting for a resource before it runs.
In such cases, it recommends using the waiters feature to help improve efficiency.

\item[Detect missing \lstinline{None} check on cached response metadata:]
Response metadata represents additional metadata included 
with a response from AWS. Response metadata varies by service, 
but all services return an AWS request ID that can be used 
in the event a service call isn't working as expected. 
If the code attempts to access the response metadata, 
\lstinline{ResponseMetadata}, without performing a \lstinline{None} 
check on the response object, then this might cause a \lstinline{NoneType} error.
To prevent this, our rule recommends adding a \lstinline{None} check
on the response object before accessing the response metadata.

\item[Detect failed records in Kinesis \lstinline{PutRecords}:]
The \lstinline{put_records} operation in AWS Kinesis service might fail, 
thereby causing loss of records. This rule detects if the code
handles the failed records from the \lstinline{put_records} operation.
In the absence of such handling of failed records, the rule recommends checking
the \lstinline{FailedRecordCount} in the \lstinline{put_records} response 
to see if there are failed records in the response. A failed record 
includes \lstinline{ErrorCode} and \lstinline{ErrorMessage} values. If 
failed records are found, the rule recommends adding them into the
next request.

\item[Detect deprecated APIs:]
This rule detects usage of deprecated APIs in Python application code.
A total of 107 deprecated API specifications are extracted from Boto3.  
The deprecated APIs are identified from the use of \lstinline{deprecated} 
trait in the API models. These API specifications are fed into the rule 
for detecting deprecated APIs in real world Python code. 

\item[Detect inefficient/redundant API chains:]
The rule for inefficient/redundant API chains detects usage of
less performant APIs or outdated APIs, an API call chain that could be
replaced with a single API call, a manual pagination operation where
the SDK provide a Paginator API to automatically perform the pagination, and 
much more. 

\item[Detect expensive client object construction in Lambda handler:]
This rule detects a Boto3 client that is initialized from a Lambda handler.
In order to speed up Boto3 client initialization and minimize the operational 
cost of the Lambda function, the rule recommends creating the client at the
level of the module that contains the handler, and then reusing it between
invocations. This is stated in the best practices for the lambda 
handler.\footnote{\url{https://docs.aws.amazon.com/lambda/latest/dg/best-practices.html\#function-code)}}

\end{itemize}

\subsection{Engagement with the AWS SDK Team}

Beyond their core value in alerting developers to bugs and improvement opportunities, the AWS best practices rules 
enable unique collaboration between the CodeGuru and AWS SDK teams. 

From our side, we provide frequent feedback to the SDK
team. Examples include developer input that indicates a missing API feature or an API's contract being unclear; 
pushback or confusion in response to our suggested fix for a best-practice violation; or customer demand for a new rule.

The AWS SDK team has also contacted us. A recent example is the SDK V2 pagination feature, where the SDK team wanted to
promote awareness of this feature, and as part of that (i) made sure that we have a corresponding rule, and (ii) took interest in
the rule's messaging and performance in the field.

\section{Experimental Results}~\label{exp-result}
In this section, we report on experiments to validate our approach for on-demand resolution of Python types. Our experiments are guided by the following research hypotheses:
\begin{enumerate}
	\item[\textbf{Hypothesis 1:}] Skipping type inference, instead relying solely on function names and arguments, is insufficient since that might lead to excessively many false positive detections.
	\item[\textbf{Hypothesis 2:}] The dataflow-based and Pyright-with-stub-based resolution strategies have complementary strengths.
	\item[\textbf{Hypothesis 3:}] A staged approach, combining dataflow and stubs with name-based resolution as a low-confidence fallback, is effective.
    	\item[\textbf{Hypothesis 4:}] The AWS best practices rules, running atop the staged algorithm, are sufficiently precise, efficient and actionable to provide value during code review.
\end{enumerate}

We note that beyond type inference, once a function call is confirmed to invoke a given AWS service, most of the rules are 
straightforward and do not require complex and/or interprocedural analysis to detect incorrect or suboptimal use of the AWS API\@.
There are few exceptions, where the actual rule's logic can be imprecise, but overall the correctness of type inference is a good proxy for the correctness of a rule finding.

We illustrate rule dependence on identification of the Boto3 service being invoked using the JSON snippet below, which is taken from our service's production configuration. The 
``Missing Pagination'' rule, whose specification is described in the snippet, searches for paginated functions
like \lstinline{list_dataset_groups} in the specific context of the \lstinline{forecast} AWS service.
Recall that these API specifications are automatically extracted from the API models in Boto3. 

\begin{lstlisting}
{
  "expectedPaginationMethods": [
    "IsTruncated",
    "NextToken"
  ],
  "paginatedMethod": "list_dataset_groups",
  "resultKeys": [
    "DatasetGroups"
  ],
  "serviceId": "forecast"
}
\end{lstlisting}

\subsection{Dataset} 

We have evaluated the strategies described in \cref{inference} using a
dataset consisting of 3,027 public GitHub repositories.
These repositories were selected based on the following criteria:
\begin{itemize}
	\item The repository contains Python source files (at least 3, and with a total of at least 100 lines of code).
	\item The repository has an MIT or Apache license.
	\item The repository has a rating of 3 stars or more.
	\item The repository makes use of the AWS SDK.
\end{itemize}

We note that the overall number of repositories meeting the first three criteria is 86,000. Approximately 3.5\% of 
these repositories either use or import the Boto3 library.

\subsection{Performance of Resolution Strategies in Isolation}

To examine the first two hypotheses laid out above, we begin by computing precision and recall for the different type resolution strategies in isolation. 
Precision is measured as the proportion of correct (TP) versus incorrect (FP) type resolutions, and recall
is measured as the proportion of correct (TP) versus missed (FN) type resolutions.
In what follows, we use the notation $t[s]$ to refer to the type of SDK service client $s$. 

\subsubsection{Type-Resolution Strategies}\label{Se:individualStrategies} 

We consider 3 different strategies for resolution of $t[s]$:
\begin{enumerate}
  \item[\emph{Strategy 1:}] Use Pyright's type inference in conjunction with
  third-party Boto3 type stubs. This strategy potentially recover types beyond the boundaries of a single function.
  \item[\emph{Strategy 2:}] Use interprocedural dataflow analysis, combining backward and forward queries.
  \item[\emph{Strategy 3:}] Match against the API name without attempting to resolve the type of the receiver,
  which is an over-approximate yet cheap approach.
\end{enumerate}
 
\subsubsection{Results} 

\Cref{result} shows the number of resolutions due to each of the strategies when applied to the GitHub dataset.
To gain qualitative insight into the results, and how many of the type resolutions are
accurate, we manually reviewed 50 Boto3 client detections, selected at random, 
for each of the three strategies for a total of 150 detections. Reviewers
consisted of five senior engineers and scientists, all expert users of the Boto3 library.

\begin{table}[t]
   \centering
   \begin{filecontents*}{result.csv}
Description,Confidence,Type Resolution Count,Precision
Pyright with Boto3 type stubs,1.0,2293,1
Dataflow tracking,1.0,3065,1
API name based resolution,0.5,5403,0.54
\end{filecontents*}
\pgfplotstabletypeset[
   columns={Description,Confidence,Type Resolution Count,Precision},
   columns/Type Resolution Count/.style={
     column type={S[table-format=4]},
     numeric as string type,
   },
   create on use/Strategy/.style={ create col/expr={ \pgfplotstablerow + 1 }, },
   ]{result.csv}
 \caption{Number of type resolutions due to each of the resolution strategies.}
 \label{result}
 \end{table}

Our qualitative analysis suggests that strategies 1 and 2 are highly precise, 
as reported in the ``Precision'' column of \cref{result}. All 50 cases
sampled for manual review were judged as correct. By contrast, for strategy 3, only 54\% of the samples (27 out of 50) were correct.
By definition, strategy 3 achieves 100\% recall and thus establishes an upper bound on the number of false negatives
due to strategies 1 and 2. 

The set of detections obtained from strategy 1 and strategy 2 are not exactly the same,
and they do not subsume each other: some strategy-1 detections are omitted by strategy 2,
and vice versa. Out of 27 true positive detections from strategy 3,
19 detections are also obtained from strategy 1 and strategy 2 combined. The remaining 8 detections (~30\%) are
exclusive to strategy 3.

\subsubsection{Discussion}

We consider the pros and cons of the three strategies in light of these results.

Strategy 1 uses third-party Boto3 type stubs, together with Pyright's type inference
to resolve AWS SDK clients. Unlike strategy 2, where type resolution occurs \emph{during} rule evaluation,
strategy 1's Pyright-derived types are available \emph{before} rule evaluation, during MU graph construction.
This allows type resolution to run once rather than on every application of every rule: a major performance boost.

On the negative side, strategy 1 suffers from low recall, as shown in
the ``Type Resolution Count'' column of \cref{result}. This is due
to the different ways in which AWS SDK clients are obtained, and in particular, the common 
case of passing them as function parameters. Pyright does not search for callers of the function,
thus assigning \lstinline{Any} as the type of the parameters unless annotations are explicitly provided.

Moving to strategy 2, the ability to perform backward dataflow tracking addresses the challenge
of passing AWS SDK clients as function parameters. Duplication of work on type resolution is mitigated
by a staged algorithm that first attempts intraprocedural resolution, then performs tracking at the file level, and finally 
at the level of the entire codebase. From our experience, and performance measurements, the staged
algorithm is quite effective. Like strategy 1, strategy 2 retains full precision, yet has much
higher recall as shown in the ``Type Resolution Count'' column of \cref{result}.

In spite of its overall effectiveness, strategy 2 --- which tracks dataflow through local variables --- can miss cases where 
the client is stored as a field or global variable. These cases are handled by strategy 1.

Our analysis of the gaps between strategies 1 and 2 is confirmed experimentally. In line with hypothesis 2,
we have found 60 detections that are exclusive to strategy 1 and 832 detections that are exclusive to 
strategy 2.

Finally, the low precision of strategy 3 (just over 50\%) confirms hypothesis 1. At the same time, 
the computational cost of strategy 3 is virtually zero, and thanks to its simplicity, it is able to sometimes
completely bypass complex tracking scenarios that are beyond the power of strategies 1 and 2. An example
is given in \cref{strategy3}, where
neither strategy 1 nor strategy 2 is able to recognize that \lstinline{self._ec2_client}
is a Boto3 client in the body of the \lstinline{ec2_client.describe_snapshots(**kwargs)} method.
Strategy 3 succeeds here simply by recognizing \lstinline{describe_snapshots}
as the name of an AWS SDK client API method.

To make use of strategy 3 in spite of its approximate nature, we ``penalize'' detections due to this strategy by assigning a confidence 
score of 0.5 to those detections compared to 1.0 if the detection is due to strategies 1 or 2,
as shown in the ``Confidence'' column of \cref{result}.
The exact value of 0.5 is arbitrary, but serves to distinguish the lower-confidence
detections of strategy 3 from the higher-confidence detections of strategies 1 or 2.
This is in line with our earlier comment that the correctness of type resolution is a 
good proxy for the correctness of a detection.

\begin{figure}
\begin{lstlisting}
class AwsClient(object):
    def __init__(self, *args, **kwargs):
        self._boto3client = None
        super(AwsClient, self).__init__(*args, **kwargs)

    def create_ec2_client(self, context=None):
        #--> (method) create_ec2_client: 
             (self: Self@AwsClient, context=None) -> Any
        access_key = CONF.aws.aws_access_key
        secret_key = CONF.aws.aws_secret_key
        region_name = CONF.aws.aws_region
        kwargs = {}
        kwargs['aws_access_key_id'] = access_key
        kwargs['aws_secret_access_key'] = secret_key
        kwargs['region_name'] = region_name

        return boto3.client('ec2', **kwargs)

    def get_aws_client(self, context):
        if not self._boto3client:
            try:
                ec2_client = self.create_ec2_client(context)
                #--> (variable) ec2_client: Any
                self._boto3client = AwsClientPlugin(ec2_client)
                #--> (variable) _boto3client: Any
            except Exception as e:
                LOG.error(_LE('Create aws client failed: \%s'), e)
                raise exception_ex.OsAwsConnectFailed

        return self._boto3client

    def describe_snapshots(self, **kwargs):
        response = self._ec2_client.describe_snapshots(**kwargs)
        #--> (variable) _ec2_client: Any
        snapshots = response.get('Snapshots', [])
        return snapshots
\end{lstlisting}
\caption{Detections from Strategy 3 that strategies 1 and 2 miss}
\label{strategy3}
\end{figure}

\subsection{Performance of Combined Resolution Strategies}

The results in \cref{Se:individualStrategies} suggest that there is benefit in combining the different strategies in light of their complementary strengths.
Starting from this motivation, we report here on experiments with ``hybrid'' resolution strategies, which we refer to as \emph{configurations}. 

\subsubsection{Type Resolution Configurations}\label{Sec:type-resolution-configs}

We consider two configurations:

\begin{description}
  \item[High Confidence (HC)] runs strategy 1, then strategy 2 where needed to complement strategy 1.
  \item[Mixed Confidence (MC)] runs strategies 1 and 2 in the same fashion as HC, but rather than giving up if both fail, proceeds to strategy 3 in an attempt to generate a low-confidence detection.
\end{description}

CodeGuru uses the confidence score to rank the detections as per the
``Confidence'' column in \cref{result}. Detections from strategy 1 and
strategy 2 rank higher than detections from strategy 3 thanks to their higher confidence score.
CodeGuru imposes different restrictions and limitations on detectors, in particular with regard to the
overall number of detections, which means that in the presence of sufficiently many high-confidence detections,
low-confidence detections are suppressed. By implication, low-confidence MC detections are not always
reported to the user.

\subsubsection{Results}

\Cref{config} reports results for both configurations, running against
the dataset of 3,027 GitHub repositories. The total
time for running each configuration is close to 5 hours.   

\begin{table}[t]
  \centering
  \newcounter{configuration}
  \begin{filecontents*}{config.csv}
Configuration,Strategies,Description,Number of Detections
HC,"1, 2",Pyright with stubs followed by dataflow,3125
MC,"1, 2, 3",All layers,5403
\end{filecontents*}
\pgfplotstabletypeset[
    columns={Configuration, Strategies, Description, Number of Detections},
    columns/Configuration/.style={ column type=c },
    columns/Number of Detections/.style={
      column type={S[table-format=4]},
      numeric as string type,
    },
    text column,
    text indicator=",
  ]{config.csv}
\caption{Type Inference Configurations}
\label{config}
\end{table}

In line with hypothesis 2, the HC configuration generates more detections than strategies 1 or 2 in isolation.
The total number of detections due to the HC configuration is 60 more than strategy 2:
exactly the number of detections that are exclusive to strategy 1.

Moving to the MC configuration, the number of detections that it generates is identical to strategy 3 in isolation,
which is expected. The important difference, however, is that most (that is, 3,125) of the detections have 
high confidence, with only 2,278 detections relying on strategy 3.

Projecting from the detections we sampled and triaged, we estimate that the MC configuration has a precision score of 0.85 along
with perfect recall, whereas the HC configuration has perfect precision but a recall score of roughly 0.72 (with the assumption that
54\% of the findings found by MC but not HC are true positives). This analysis supports hypothesis 3, which favors use
of strategy 3 as part of the combined strategy rather than relying only on the high-confidence strategies.

\subsection{Real-world Feedback on the Rules}

Beyond our offline study, we also report on data from the field driven by comments that CodeGuru has left on code reviews in production. In this use case, CodeGuru posts comments on code reviews just as a human reviewer would. We have augmented the comment UI with a feedback menu, so that a developer can optionally rate a detection as ``Useful'', ``Not Useful'' or ``Not Sure'' and/or provide free-form textual feedback. These feedback mechanisms give the CodeGuru team insight into the performance of different detectors and enable detector tuning over time.

For AWS best practices, each CodeGuru comment contains two key fields:

\begin{enumerate}
\item One or two paragraphs explain what the issue is, and why fixing it is important. For example, in the case of a batch operation whose output is ignored, the explanation states that even if some items are not processed successfully, the batch operation might still complete successfully without raising an exception.

\item A ``Learn More'' hyperlink directs the user to the appropriate section in the Boto3 online documentation for complete information on the API in question.
\end{enumerate}
 


We provide lower-bound metrics to give a sense of the size of CodeGuru's input funnel.
In the studied time period of 10 weeks, CodeGuru analyzed $\gg1,000,000$ lines of code.
We applied $\gg10$ detectors, yielding $\gg10,000$
AWS best practice recommendations, which we reported to $\gg 1,000$
developers.

We note that by definition, the codebases involved in this study are all live (undergoing code reviews and modifications). These are Python cloud services and applications that make use of Boto3, where the developers are industry practitioners with Python and cloud background. Hence we assign high weight to their feedback on CodeGuru detections.

In CodeGuru, we measure \emph{acceptance} as an indication of whether or not developers have found a given rule's review comments
useful. Given a set of ``Useful'' ($U$), ``Not Useful'' ($\mathit{NU}$) and ``Not Sure'' ($\mathit{NS}$) ratings, we compute
acceptance as the ratio
\begin{equation*}
  \frac{|U|}{|U| + |\mathit{NU}| + |\mathit{NS}|}
\end{equation*}
where by $|U|$ we mean the number of ``Useful'' feedback points, and analogously
for $\mathit{NU}$ and $\mathit{NS}$. Note, importantly, that we conservatively treat ``Not Sure'' the same as ``Not Useful''.

\Cref{acceptance} shows the acceptance data for eight of the Python AWS best practices rules for a time period 
of 10 weeks.
We obtained $\gg100$ feedback points from a population of $\gg100$ developers through the feedback UI described above.
As reported in \cref{acceptance}, developers accepted over 85\% of the recommendations
made by five out of the eight rules, and almost 83\% of the overall recommendations.
\begin{table}[t]
  \centering
  \begin{filecontents*}{acceptance.csv}
Rule,Acceptance Rate
Detect missing Pagination,.7504
Data loss in Batch APIs,1
Use Waiters instead of Polling APIs,.52
Detect failed Records in Kinesis PutRecords,1
Detect deprecated APIs,.889
Detect usage of inefficient/redundant API chains,.857
Missing None check on cached response metadata,.857
Detect expensive client object construction in Lambda handler,.758
\end{filecontents*}
\pgfplotstabletypeset[
    columns/Rule/.style=text column,
  ]{acceptance.csv}
  \caption{Acceptance rate per rule from developer feedback during code review}
  \label{acceptance}
\end{table}

\begin{table}[t]
  \centering
  \begin{filecontents*}{configuration.csv}
Detection Group,High Confidence Detections Proportion,Low Confidence Detections Proportion
All,.88,.115 
Accepted,.93,.07
Not Accepted,.84,.16
\end{filecontents*}
\pgfplotstabletypeset[
    columns/Detection Group/.style=text column,
  ]{configuration.csv}
  \caption{Breakdown of the detections from \cref{acceptance} by confidence level}
  \label{configuration}
\end{table}

Only one of the eight rules, ``Use Waiters instead of Polling APIs'', has an acceptance rate below 75\%.
Our analysis of this rule's performance, including communication with some of the developers who left
feedback on its detections, suggests that the gap between acceptance and correctness is important. Developers
often acknowledge the detection as correct, but push back for one or more of the following reasons:
\begin{itemize}
  \item The intent of the PR is different, and they prefer not to merge multiple unrelated changes into the same PR\@.
  \item The change is applicable, but requires upgrading the codebase to use the latest AWS Python SDK, which again
  exceeds the scope of the PR\@.
  \item The change is not applicable, since the code in question is test code or there is no concern about polling in the
  given context.
\end{itemize}
It is worth adding that outside the time period reported here, we have seen 
multiple weeks where acceptance rate for ``Use Waiters instead of Polling APIs'' was high.

Overall, acceptance data from the field supports hypothesis 3 in showing that developers mostly find the detections by
to the Python AWS best practices rules useful. These are made using the MC configuration, which integrates all three of
the resolution strategies described in \cref{Se:individualStrategies}.

From our conversations with developers, the textual feedback they provided, and our own review of some of the detections
and their corresponding feedback, we have identified two main factors that contribute to the usefulness of our rules:
\begin{itemize}
\item Missed features: SDK changes across versions, in particular new features, 
are sometimes missed by developers. Pagination, retry and error handling are examples of such features,
where developers not familiar with these built-in capabilities sometimes implement ``manual'' mechanisms instead.
Another example is manual polling versus the recommended use of the waiter utility.
\item Missed expectations: Developers sometimes assume, rather than verify, the functionality
of a given API or the role of a given parameter. An example is the \lstinline{QueryResponse::hasItems} method,
whose (boolean) return value is sometimes incorrectly interpreted to mean that the response 
contains a non-empty collection of items, where what is in fact meant is that response defines an \lstinline{Items} property.
To make sure whether any items are contained in the response, the developer needs to also check 
\lstinline{Items::isEmpty}. Mistakes like this can lead to large-scale operational failures.
\end{itemize}

\Cref{configuration} reports the breakdown, by confidence level (high versus low), for the detections in \cref{acceptance}.
In sharp contrast to the distribution due to strategy 3 from the offline study, where approximately 45\% of the detections had
a low confidence score, the hybrid inference strategy leans heavily towards high-confidence detections (88\% of all detections). 
This is consistent with the suppression policy described above, in \cref{Sec:type-resolution-configs}, for low-confidence detections.
The tradeoff that the hybrid strategy offers in the presence of confidence-based suppression is appealing, in that 
low-confidence detections are typically shadowed by high-confidence detections, which 
limits the impact of such detections on precision and allows them to play an important role in pushing coverage upwards when 
high-confidence detections are absent.
Also note, from \Cref{configuration}, that the proportion of low-confidence detections among ``Not Accepted'' detections is higher compared to
``Accepted'' detections (16\% versus 7\%), which is consistent with the data from the offline study. 

Overall, our analysis of detections from the field, and how these map back to the hybrid strategy, are in support of hypothesis 4.
Developers tend to view our AWS best practices recommendations as useful. Most of the recommendations build on
high-confidence type inference, with some remaining cases benefitting from the low-confidence resolution strategy.

\section{Conclusion and Future Work}\label{Sec:conclusion}
We have presented an industrial-strength framework for precise static analysis 
of Python applications that use AWS cloud services.  In support of this
goal, we have developed
a novel type inference system for identifying and tracking AWS service clients 
in real-world Python applications.
Our Python MU graph IR is suitable for building a wide range of static analyses or
best-practice rules for Python applications.
Furthermore, the Golden Query Language provides the right level of abstraction with its encapsulation, optimization and reuse features to  
develop static analysis rules that can be evaluated at different scopes, from single functions to entire applications.

Experiments on 86K open-source Python GitHub repositories show that 
individual inference strategies have complementary strengths.  The most 
effective solution, then, is a layered approach that combines Pyright with Boto3 stubs, 
custom dataflow analysis in GQL, and name-based resolution as a low-confidence fallback.
Our layered strategy achieves 85\% precision and 100\% recall in typing
relevant Boto3 values in Python client code.
The ultimate authorities on the value of our approach are real-world
developers, with no ties to the authors.  Those developers accepted more
than 85\% of the recommendations made by five out of eight rules, and roughly 83\%
of the recommendations on average.

In the future, we plan to extend and generalize our type inference infrastructure to 
other rule suites and properties that apply to Python programs. We are also examining 
ways to reuse our work on Python on-demand type inference when adding support for 
other languages sans static typing.

\appendix

\section{Additional GQL Operations}\label{Sec:MoreGQLOps}

\Cref{Sec:Operations} walked the reader through four categories of GQL operations: core, filter, transform, and second-order operations.
We revisit these categories here, and provide additional representative examples of operations from each of these categories. These are not exhaustive,
and are shared with the purpose of conveying a more complete illustration of the capabilities that GQL offers.
 
\subsection{Representative Core Operations}\label{Sec:rep-core-ops}

\Cref{GQLCoreOperations} lists a representative subset of the GQL core operations. Some have already been explained. We explain a few more below.

\begin{table}
\begin{GqlApi}
\text{\lstinline{as}} & Stores the current match frontier as the specified variable. \\
\text{\lstinline{check}} & Transitions from precondition to postcondition evaluation. \\
\text{\lstinline{reset}} & Resets the match frontier to all graph nodes. \\
\text{\lstinline{withAuxiliaryState}} & Sets arbitrary mutable state to be consulted, and manipulated, during rule evaluation. \\
\text{\lstinline{withComment}} & Sets the rule's comment (or description). \\
\text{\lstinline{withCommentOverride}} & Overrides the current rule comment while optionally taking the current evaluation context into account. \\
\text{\lstinline{withGraphics}} & Outputs a visual representation of the target code that highlights matched nodes (useful for debugging purposes). \\
\text{\lstinline{withId}} & Loads the node set mapped to the provided id. \\
\text{\lstinline{withInstrumentation}} & Enables (read-only) inspection of the current match frontier, for example to debug a rule or compute metrics. \\
\text{\lstinline{withName}} & Sets the rule's name. \\
\end{GqlApi}
\caption{Representative collection of GQL core operations}
\label{GQLCoreOperations}
\end{table}

\lstinline{withInstrumentation} allows access to intermediate match frontiers throughout rule evaluation. Consider for example the below rule snippet:
\begin{lstlisting}[language=Java]
.withMethodCallFilter("foo")
.withInstrumentation(mr -> System.out.println(mr))
\end{lstlisting}
Here \lstinline{mr} denotes the match frontier post the function-call filter. Given a function that contains one or more such calls, those matches would be printed using \lstinline{println}.
This is a simple illustration of how \lstinline{withInstrumentation} can help with debugging tasks, or other use cases such as profiling or analytics.

Another handy construct is \lstinline{reset}. A common use for this construct is sequencing of independent operations. Here is an example:
\begin{lstlisting}[language=Java]
.withMethodCallFilter("foo")
.as("fooCalls")
.reset()
.withMethodCallFilter("bar")
\end{lstlisting}
The rule first scans for ``\lstinline{foo}'' calls and stores these as variable \lstinline{"fooCalls"}, then performs an unrelated search for \lstinline{"bar"} calls.

\subsection{Representative Filter Operations}\label{Sec:rep-filter-ops}

\Cref{GQLFilterOperations} describes representative GQL filter operations.

Note that some of these operations operate on
IDs, as managed through the \lstinline{as} and \lstinline{withId} constructs. As an example, \lstinline{withReceiverByIdFilter} takes the ID of a stored
match frontier as input, and allows through incoming function calls that have as their receiver a node belonging in that match frontier. 

We also note that the filters perform analysis of varying depth behind the scenes. \lstinline{withActionFilter} is an example of a basic filter that 
simply pattern matches against action nodes in the incoming match frontier, whereas \lstinline{withConstantArgumentFilter} is an advanced filter
that performs constant propagation, and even a limited form of string analysis, to track and compute constant values.

\begin{longtblr}[caption={Representative collection of GQL filter operations}, label=GQLFilterOperations]{colspec=lX, rows={rowsep=.125\baselineskip}, rowhead=1}
\toprule
Builder API & Description \\
\midrule
\text{\lstinline{withActionFilter}} & Accepts actions matching the provided regex. \\
\text{\lstinline{withArgumentByTypeFilter}} & Accepts action and control nodes with an argument whose type matches the provided regex. \\
\text{\lstinline{withArgumentContextFilter}} &  Accepts action nodes having an argument defined in the provided syntactic context (for example, loop, conditional or switch statement). \\
\text{\lstinline{withArgumentValueFilter}} & Accepts action nodes with a matching argument (multiple overloads). \\
\text{\lstinline{withCatchClauseFilter}} &  Accepts action nodes with a corresponding catch clause. \\
\text{\lstinline{withCaughtExceptionFilter}} & Accepts action nodes with a target catch clause whose exception matches the provided regex. \\
\text{\lstinline{withConstantArgumentFilter}} & Accepts action nodes with a constant argument matching the provided regex. \\
\text{\lstinline{withContextFilter}} & Accepts nodes that reside within the provided control context. \\
\text{\lstinline{withControlUserFilter}} & Accepts nodes used by the specified control constructs (like do, while, and so on). \\
\text{\lstinline{withDataByNameFilter}} & Accepts data nodes whose name matches the provided regex. \\
\text{\lstinline{withDataByTypeFilter}} & Accepts data nodes whose type matches the provided regex. \\
\text{\lstinline{withDataFromIdFilter}} & Accepts nodes that receive data (transitively) from nodes mapped to the provided id. \\
\text{\lstinline{withDataFromParameterFilter}} &  Accepts nodes that receive data (transitively) from function parameters. \\
\text{\lstinline{withDataFromResultFilter}} & Accepts nodes that receive data (transitively) from a node (possibly) returned by the function. \\
\text{\lstinline{withDeclaringTypeFilter}} & Accepts function calls whose enclosing type matches the provided regex. \\
\text{\lstinline{withDownstreamConditionalCheckFilter}} & Accepts nodes whose data flows (transitively) into a conditional check (including loop tests). \\
\text{\lstinline{withLocalDataFromIdFilter}} & Same as above, but with the restriction to local data flow (blocking flow across function calls). \\
\text{\lstinline{withLowerCaseArgumentFilter}} & Accepts action nodes with a constant lower-case string argument. \\
\text{\lstinline{withMethodCallFilter}} & Accepts function calls matching the provided regex. \\
\text{\lstinline{withMethodCallGroupFilter}} & Accepts a group of calls if their names, and how they relate to each other, meet the spec (for example, if they share the same receiver or are ordered as specified). \\
\text{\lstinline{withNodeByTypeFilter}} & Accepts nodes whose corresponding AST node type matches the provided specification (for example, \text{\lstinline{ARRAY_ACCESS}} or \text{\lstinline{CAST_EXPRESSION}}). \\
\text{\lstinline{withNumberOfArgumentsFilter}} & Accepts action nodes with the provided number of arguments. \\
\text{\lstinline{withReceiverByIdFilter}} & Accepts action nodes whose receiver is mapped to the provided id. \\
\text{\lstinline{withReceiverByTypeFilter}} & Accepts action nodes with a receiver whose type matches the provided regex. \\
\text{\lstinline{withReturnValueFilter}} & Accepts data nodes that are (possibly) returned by the function. \\
\text{\lstinline{withUpperCaseArgumentFilter}} &  Accepts action nodes with a constant upper-case string argument. \\
\text{\lstinline{withUserFilter}} &  Accepts data nodes that have users matching the provided regex. \\
\bottomrule
\end{longtblr}

\subsection{Representative Transform Operations}\label{Sec:rep-transform-ops}

\Cref{GQLTransformOperations} presents a small subset of the GQL transform operations. Notice in particular the
data-flow and taint constructs. These are used extensively across GQL rules, especially in the security category, and enable 
advanced tracking features (like rule-specific combinations of forward and backward tracking, or use of different taint configurations
within and across rules).

\begin{table}
\begin{GqlApi}
\text{\lstinline{withArgumentsTransform}} & Transforms an action node to its respective arguments. \\
\text{\lstinline{withCatchClauseTransform}} & Transforms an action node to its target catch nodes. \\
\text{\lstinline{withControlDependenciesTransform}} & Transforms a node to its set of control dependencies. \\
\text{\lstinline{withDataDependenciesTransform}} & Transforms a node to its set of (transitive) data dependencies. \\
\text{\lstinline{withDataDependentsTransform}} & Transforms a node to the set of nodes that are data dependent on it. \\
\text{\lstinline{withDefinitionTransform}} & Transforms data nodes to their defining action node (if exists). \\
\text{\lstinline{withTaintFlowFromTransform}} & Transforms a node treated as a ``sink'' to the ``source'' nodes that reach it, where the taint specification is provided through this API's arguments. \\
\text{\lstinline{withTaintFlowToTransform}} & Transforms a node treated as a ``source'' to the ``sink'' nodes that it reaches, where the taint specification is provided through this API's arguments. \\
\text{\lstinline{withReceiverTransform}} & Transforms an action node with a receiver to the receiver, or else $\emptyset$ for actions sans receiver. \\
\text{\lstinline{withThenTransform}} & Transforms nodes to the nodes that occur after them per the control-flow relation. \\
\text{\lstinline{withUsersTransform}} & Transforms a data node to its users. \\
\end{GqlApi}
\caption{Representative collection of GQL transform operations}
\label{GQLTransformOperations} 
\end{table}

\subsection{Representative Second-order Operations}\label{Sec:rep-2ndorder-ops}

\Cref{GQLSecondOrderOperations} shows a subset of GQL's second-order operations. We discuss several additional constructs beyond those already described.

\begin{table}
\begin{GqlApi}
\text{\lstinline{withAllOf}} & Evaluates to a relational mapping from subrules to their respective result node sets if all are successful, or else $\emptyset$. \\
\text{\lstinline{withAdditional}} & Evaluates to the union of the result due to the subrule provided as argument and the incoming match frontier. \\
\text{\lstinline{withAllOf}} & Evaluates to a relational mapping from subrules to their respective result node sets if all evaluate successfully, or else $\emptyset$. \\
\text{\lstinline{withAnyOf}} & Evaluates to a relational mapping from subrules to their respective result node sets. \\
\text{\lstinline{withClosure}} & Evaluates to the result of the subrule provided as argument, which is the recommended way to integrate helper functions into a GQL rule. \\
\text{\lstinline{withIndependent}} & Evaluates to the result of the subrule provided as argument. \\
\text{\lstinline{withInterproceduralMatch}} & Applies the provided subrule to an interprocedural scope. The scope, and how to traverse it, are specified by the user. \\
\text{\lstinline{withLanguageSpecific}} & Contains language-specific subrules. The appropriate subrule is selected per the program's source language. \\
\text{\lstinline{withNegationOf}} & Evaluates to the negation of the subrule provided as argument: $\emptyset$ if the subrule has a non-empty result, or else the input node set. \\
\text{\lstinline{withOneOf}} & Evaluates to the first successful subrule if exists, or else evaluates to $\emptyset$. \\
\end{GqlApi}
\caption{Representative collection of GQL second-order operations}
\label{GQLSecondOrderOperations} 
\end{table}

The \lstinline{withNegationOf} construct accepts a single subrule, and checks for the negation of its evaluation. That is, \lstinline{withNegationOperation} evaluates to the input match frontier if the subrule evaluates to $\emptyset$, or else it evaluates to $\emptyset$. Here is an example:
\begin{lstlisting}[language=Java]
.withNegationOf(b -> b.withMethodCallFilter("foo"))
\end{lstlisting}
If the incoming frontier contains a \lstinline{foo} call, then the negation operation fails. Otherwise, the incoming frontier is propagated through \lstinline{withNegationOf}.

A useful construct in the case of multi-language rules is \lstinline{withLanguageSpecific}. Loosely, this construct acts like a \lstinline{switch} statement, enabling the user to provide per-language subrules to implement certain rule steps (whereas other steps are shared across different languages). If for example the rule enforces correct use of a paginated DynamoDB API, the first step being to ensure that the call is against a DynamoDB client, then \lstinline{withLanguageSpecific} helps with this task by allowing the user to check (i) the type of the receiver in the case of Java, and (ii) the definition of the receiver --- using our staged type inference algorithm --- in the case of Python.

%

\bibliographystyle{plainurl}
\bibliography{dblp, miscellaneous, web-sites}

\end{document}